\documentclass[10 pt, aps, pre,twocolumn]{revtex4-1}

\usepackage{epsfig} 
\usepackage{amsmath} 
\usepackage{amssymb} 
\usepackage{amsfonts}
\usepackage{bm}
\usepackage{graphicx}
\usepackage[normalem]{ulem}
\usepackage[usenames,dvipsnames,table]{xcolor}

\def\revision{\textbf}

\def\bx{\bm{x}}

\def\br{\bm{r}}
\def\bR{\bm{R}}
\def\R{\mathbb{R}}
\def\>{\rightarrow}
\def\({\left(}
\def\){\right)}

\def\>{\rightarrow}
\def\grad{\nabla}

\newcommand{\mb}[1]{\bm{{#1}}}
\newcommand{\mc}[1]{\mathcal{{#1}}}
\newcommand{\norm}[1]{\left\|\, #1 \,\right\|}

\begin{document}

\title{Collective motion patterns of swarms with delay coupling: theory and experiment}
\author{Klementyna Szwaykowska}
\email{klementyna.szwaykowska.ctr@nrl.navy.mil}
\affiliation{U.S. Naval Research Laboratory\\Code 6792\\ Plasma Physics Division\\ Nonlinear
Dynamical Systems Section\\ Washington, DC}
\author{Luis Mier-y-Teran Romero}
\affiliation{Johns Hopkins University\\
Bloomberg School of Public Health\\
Department of Epidemiology\\
Baltimore, MD}
\author{Christoffer R. Heckman}
\affiliation{Department of Computer Science\\
University of Colorado\\
Boulder, CO}
\author{Dan Mox}
\affiliation{Drexel University\\
Mechanical Engineering \& Mechanics Department \\
Scalable Autonomous Systems Laboratory\\
Philadelphia, PA}
\author{M. Ani Hsieh}
\affiliation{Drexel University\\
Mechanical Engineering \& Mechanics Department \\
Scalable Autonomous Systems Laboratory\\
Philadelphia, PA}
\author{Ira B. Schwartz}
\email{ira.schwartz@nrl.navy.mil}
\affiliation{U.S. Naval Research Laboratory\\Code 6792\\ Plasma Physics Division\\ Nonlinear
Dynamical Systems Section\\ Washington, DC}
\date{\today}


\begin{abstract}

The formation of coherent patterns in swarms of interacting self-propelled autonomous agents is a subject of great
interest in a wide range of application areas, ranging from engineering and physics to biology. In this paper, we model
and experimentally realize a mixed-reality large-scale swarm of delay-coupled agents. The coupling term is modeled as a
delayed communication relay of position. Our analyses, assuming agents communicating over an Erd{\"o}s-Renyi network,
demonstrate the existence of stable coherent patterns that can only be achieved with delay coupling and that are robust
to decreasing network connectivity and heterogeneity in agent dynamics. We also show how the bifurcation structure for
emergence of different patterns changes with heterogeneity in agent acceleration capabilities and limited connectivity
in the network as a function of coupling strength and delay. Our results are verified through simulation as well as
preliminary experimental results of delay-induced pattern formation in a mixed-reality swarm.

\end{abstract}


\maketitle

\section{introduction}

The emergence of complex dynamical behaviors from simple local interactions between pairs of agents
in a group is a widespread phenomenon over a range of application domains. Many striking examples
can be found in biological systems, from the microscopic (e.g., aggregates of bacterial cells or the
collective motion of skin cells in wound healing) \cite{Budrene1995,Polezhaev2006,Lee2013} to
large-scale aggregates of fish, birds, and even humans \cite{Tunstrom2013,Helbing1995,Lee2006}.
These systems are particularly interesting to the robotics community because they allow simple
individual agents to achieve complex tasks in ways that are scalable, extensible, and robust to
failures of individual agents. In addition, these aggregate behaviors are able to form and persist
in spite of complicating factors such as communication delay and restrictions on the number of
neighbors each agent is able to interact with, heterogeneity in agent dynamics, and environmental
noise. These factors, and their effects on swarm behaviors, are the focus of our current work.

A number of studies show that even with simple interaction protocols, swarms of agents are able to
converge to organized, coherent behaviors. Existing literature on the subject provides a wide
selection of both agent-based \cite{Helbing1995,Lee2006,Vicsek2006,Tunstrom2013} and continuum models
\cite{Edelstein-Keshet1998,Topaz2004,Polezhaev2006}. One of the  earliest agent-based models of
swarming is Reynolds's \textit{boids} \cite{Reynolds1987}, which simulates the motion of a group of
flocking birds. The boids follow three simple rules: collision avoidance, alignment with
neighbors, and attraction to neighbors. Since the publication of Reynolds's paper, many models based
on ``zones'' of attraction, repulsion, and/or alignment have been used as a means of realistically
modeling swarming behaviors \cite{Miller2012,Tarras2013,Viragh2014}. Systematic numerical studies of
discrete flocking based on alignment with nearest neighbors were carried out by Vicsek \textit{et al.}
\cite{Vicsek1995}. Stochastic interactions between agents are modeled in \cite{Nilsen2013}. In
recent years, improved computer vision algorithms have allowed researchers to record and analyze the
motions of individual agents in biological flocks, and formulating more accurate, empirical models
for collective motion strategies of flocking species including birds and fish
\cite{Ballerini2008,Katz2011,Calovi2014}.

Despite the multitude of available models, how group motion properties emerge from individual agent
behaviors is still an active area of research. For example, \cite{Viscido2005} presents a
simulation-based analysis of the different kinds of motion in a fish-schooling model; the authors
map phase transitions between different aggregate behaviors as a function of group size and maximum
number of neighbors that influence the motion of each fish. In \cite{Lee2006}, the authors use
simulation to study transitions in aggregate motions of prey in response to a predator attack.

Interaction delay is a ubiquitous problem in both naturally-occurring and artificial systems, including blood cell
production and coordinated flight of bats \cite{Martin2001,Bernard2004,Monk2003,Giuggioli2015,Forgoston2008}.
Communication delay can cause emergence of new collective motion patterns and lead to noise-induced switching between bistable
patterns \cite{Romero2011,Romero2012,Lindley2013a}; this, in turn, can lead to instability in robotic swarming systems
\cite{Viragh2014,Liu2003}. Thus, understanding the effects of delay is key to understanding many swarm behaviors in natural,
as well as engineered, systems.

In addition, many models make the mathematically simple but physically implausible assumption that swarms are
globally coupled (that is, each agent is influenced by the motion of all other agents in the swarm)
\cite{Motsch2011,Chen2011,Chen2014,Lee2006,Vecil2013a}. Global coupling is easier to analyze and a
reasonable assumption in cases of high-bandwidth communication, with a sufficiently small number of
agents. In contrast, we are interested in the collective motion patterns that emerge when global
communication cannot be achieved. New behaviors can unexpectedly emerge when the communication structure
of a network is altered, as in \cite{VonBrecht2013}, where the stability of solutions for compromise dynamics over an
Erd{\"o}s-Renyi communication network is considered. However, in our system, we show robustness of emergent motion
patterns to loss of communication links in presence of delayed coupling.

A third effect we consider is agent heterogeneity. Most existing work assumes that the members of the swarm are
identical. However, many
practical applications involve swarms that are composed of agents with differing dynamical
properties from the onset, or that become different over time due to malfunction or aging. Swarm
heterogeneity leads to interesting new collective dynamics such as spontaneous segregation of the
various populations within the swarm; it also has the potential to erode swarm cohesion. In biology,
for example, it has been shown that sorting behavior of different cell types during the development
of an organism can be achieved simply by introducing heterogeneity in inter-cell adhesion properties
\cite{Steinberg1963,Graner1993}. In robotic systems, allowing for heterogeneity in dynamical behaviors of swarm agents
gives greater flexibility in system design, and is therefore desirable not only from a theoretical but also from a
practical point of view.

A number of existing works on the spatio-temporal patterns of swarm dynamics present results that
are valid in the thermodynamic limit, where the number of agents is assumed to be very large
\cite{Vicsek2006,Edelstein-Keshet1998,Topaz2004,Romero2011,Mier-y-Teran-Romero2012a,Mier-y-Teran-Romero2014,Leverentz2009,Burger2013,Szwaykowska2014}.
We follow this mean-field approach to analytically predict transitions between regimes of different
collective swarm motions, as a function of model parameters, for swarms with random communication
graphs, under communication delay and agent heterogeneity.

We also run extensive numerical simulations to test the limits of the thermodynamic model, by
limiting the number of agents in the swarm. Extremely large experiments with distributed communication
architecture are difficult to run either in the  lab or field. The complex logistical
issues of deploying a swarm of even fifty autonomous fixed-wing aircraft are clearly seen in \cite{Day2015}. However,
most experimental work on multi-robot cooperative motion uses much smaller groups
\cite{Jia2014,Chen2014b} (a notable exception is \cite{Rubenstein2014}, which uses a centralized
controller to overcome the logistical issues involved with coordinating a very large group of agents).

We consider a generalized model of delay-coupled agents given in \cite{Szwaykowska2015}.
We show, through a combination of theory, simulation, and experiment, that the collective motion
patterns observed in the globally-coupled system in the presence of delayed communication
\cite{Mier-y-Teran-Romero2012a,Szwaykowska2014} persist as the degree of the communication network
decreases, though some characteristics of the motion patterns are altered.

\section{Model Formulation}

Consider a swarm of delay-coupled agents in $\R^d$. We assume $d=2$ in the remainder of this paper,
but our results may be generalized in a straightforward way for higher dimensions. Each agent is indexed by
$i \in \{1,\ldots, N\}$. We use a simple but general model for swarming motion. Each agent has a
self-propulsion force that strives to maintain motion at a preferred speed and a coupling force that
governs its interaction with other agents in the swarm. The interaction force is defined as the
negative gradient of a pairwise interaction potential $U(\cdot,\cdot)$. All agents follow the same
rules of motion; however, mechanical differences between agents may lead to heterogeneous dynamics;
this effect is captured by assigning different acceleration factors (denoted $\kappa_i$) to the
agents.

Agent-to-agent interactions occur along a graph $\mc{G} = \{\mc{V},\mc{E}\}$, where $\mc{V}$ is
the set of vertexes $v_i$ in the graph and $\mc{E}$ is the set of edges $e_{ij}$. The vertexes
correspond to individual swarm agents, and edges represent communication links; that is, agents $i$
and $j$ communicate with each other if and only if $e_{ij} \in \mc{E}$. All communications links are
assumed to be bidirectional, and all communications occur with a time delay $\tau$. Let
$\mc{N}_i = \{v_j \in \mc{V} : e_{ij} \in \mc{E}\}$ denote the set of neighbors of
agent $i$. Unlike in \cite{Szwaykowska2015}, we consider both heterogeneity in the agent accelerations
and distributed coupling in the interagent communication network. The motion of agent $i$ is governed by the following equation:
\begin{equation} \label{Eq:agenti}
 \ddot{\mb r}_i = \kappa_i(1-\norm{\dot{\mb r}_i}^2)\dot{\mb r}_i - \kappa_i \sum_{j \in \mc{N}_i}\grad_x U({\mb r}_i(t),{\mb r}_j^\tau(t)),
\end{equation}
where superscript $\tau$ is used to denote time delay, so that
${\mb r}_j^\tau(t) = {\mb r}_j(t-\tau)$, $\norm{\cdot}$ denotes the Euclidean norm, and $\grad_x$
denotes the gradient with respect to the first argument of $U$. The first term in
Eq.~\ref{Eq:agenti} governs self-propulsion.

We use mean-field dynamics in the limit as $N \> \infty$ to examine dynamical pattern formation in
the aggregate system and describe a bifurcation diagram showing transitions between different
motion patterns as we vary model parameters. We use a harmonic interaction potential with
short-range repulsion
\begin{equation} \label{eq:U}
 U(\bx_i,\bx_j^\tau) = c_r e^{-\frac{\norm{\bx_i - \bx_j}}{l_r}} + \frac{a}{2N}(\bx_i - \bx_j^\tau)^2.
\end{equation}
The harmonic potential was introduced in \cite{Mikhailov1999} and \cite{Erdmann2005} to model
interactions between agents. This choice can be justified empirically to some extent by noting
that, for example, \cite{Katz2011} measures a harmonic interaction in golden shiner fish, with an
added short-range repulsion.
In the model, the repulsion force acts over a characteristic distance determined by $l_r$. For
$c_r$ and $l_r$ sufficiently small, the repulsion force can be treated as a small perturbation of
the harmonic interaction potential. While we derive analytical results under the assumption $c_r = 0$,
our numerical studies indicate that the collective dynamics of the swarm are not significantly
altered by the introduction of short-range weak repulsion terms \cite{Romero2012}.
Furthermore, we assume that the communication network for the swarm is a fixed Erd\"{o}s-Renyi
random graph constructed at the beginning of the simulation and invariant in time.

We examine the dynamics of the system analytically in the limit where $\mc{G}$ is almost complete
($\frac{(N-1)-|\mc{N}_i|}{N-1} \ll 1$, where $|\mc{N}_i|$ is the number of neighbors of
node $i$), and show via simulations that the approximations made in the almost-complete limit hold
closely even as the mean coupling degree is reduced to less than $50\%$ of possible links, for an
Erd\"{o}s-Renyi communication network. We also show how varying network degree and heterogeneity in
the agent dynamics affects the bifurcation structure of the swarm motion patterns, and present
numerical simulation to verify that our theoretical results give a good approximation to the true swarm dynamics
even as the number of agents in reduced to as few as twenty.

\section{System Dynamics in the Mean-Field}

We start our analysis of the system dynamics by considering the mean-field motion, in the limit as $N \> \infty$. Let
${\mb R}(t) = \frac{1}{N} \sum_{i=1}^N {\mb r}_i(t)$ denote the position of the center of mass of the swarm. Then,
applying the change of variables $\delta {\mb r}_i(t) = {\mb r}_i(t) - {\mb R}(t)$ and substituting in Eq.~\ref{eq:U}
for $U$, Eq.~\ref{Eq:agenti} can be written as
\begin{equation} \label{eq:ddeltar}
\begin{split}
 \ddot{\bR} + \delta \ddot{\br}_i &= \kappa_i \left(1- \norm{\dot{\bR} + \delta \dot{\br}_i}^2\right) (\dot{\bR} + \delta \dot{\br}_i) \\
& - \frac{a \kappa_i}{N} \sum_{j \in \mc N_i} \( \bR + \delta \br_i - \bR^\tau - \delta \br_j^\tau \),
\end{split}
\end{equation}
\revision{where $\bR^\tau = \bR(t-\tau)$.}
The motion of the center of mass can be found by summing the above equations over all $i$ and dividing by $N$; after
simplifying, we have:
\begin{equation}
\begin{split}
\ddot{\mb R}
&= \langle \kappa_i \rangle \(1-\norm{\dot{\bR}}^2\) \dot{\bR} + \frac{1}{N}\(1-\norm{\dot{\bR}}^2\) \sum_{i=1}^N \kappa_i \delta \br_i \\
& - \frac{1}{N}\sum_{i=1}^N (\norm{\delta \dot{\mb r}_i}^2 + 2 \langle \dot{\bR}, \delta \dot{\br}_i \rangle) (\dot{\bR} + \delta \dot{\br}_i) \\
& - \frac{a}{N} \langle \kappa_i |\mc N_i| \rangle ({\bR} - {\bR}^\tau) - \frac{a}{N^2} \sum_{i=1}^N \kappa_i \sum_{j \in \mc N_i} \( \delta {\br}_i - \delta {\br}_j^\tau \),
\end{split}
\end{equation}
where $\langle \cdot \rangle$ is the average over $i$ and $\langle \cdot, \cdot \rangle$ is the dot product in $\R^2$.
Our previous work shows that in many instances either individual deviations from the center of mass are small, or in
aggregate they tend to cancel out over the whole population \cite{Romero2012} (we discuss situations in which this
assumption breaks in a later section). Then, neglecting all terms of order $\delta \bm{r}_i$ and in the limit
$N \> \infty$, the center of mass motion is given approximately by
\begin{equation} \label{eq:ddotR}
 \ddot{\mb R} = \langle \kappa_i \rangle (1-\|\dot{\mb R}\|^2)\dot{\mb R} - a \bar d ({\bR} - {\bR}^\tau),
\end{equation}
where $\bar d$ is the weighted, normalized mean degree of a node for $N \> \infty$, given by
\begin{equation*}
\bar d = \lim_{N \> \infty} \frac{\langle \kappa_i |\mc N_i| \rangle}{N}.
\end{equation*}

Note, in the current paper we do not assume correlations between $\kappa_i$ and $|{\mc N}_i|$ so that
$\langle \kappa_i |\mc N_i| \rangle \rightarrow \langle \kappa_i\rangle \langle|\mc N_i| \rangle$ as
$N\rightarrow \infty$.

The system dynamics are described by the set of coupled differential equations in Eq.~\ref{eq:ddeltar} and
\ref{eq:ddotR}. Note, however, that these equations are not all independent since, from the definition of $\delta \br_i$,
it follows that $\sum_{i=1}^N \delta \br_i = \sum_{i=1}^N \delta \dot{\br}_i = \sum_{i=1}^N \delta \ddot{\br}_i = 0$.

Quite remarkably, simulation results indicate that the system exhibits similar collective motions to the globally
coupled, homogeneous case, even as the variance of $\kappa_i$ is significantly varied or as $\bar d$ is significantly
decreased. These collective motions include ``translation'', where the entire swarm as a group travels along a
straight-line trajectory at constant speed; ``ring'' motion, where the swarm agents form concentric
counter-rotating rings about the stationary center of mass; and ``rotating'' motion, where the agents collapse to a
small volume and collectively rotate about a fixed point. The collective motions of the swarm and the effects of
non-global coupling and heterogeneity are described in more detail in the following sections.

\section{Collective swarm motions}

The quasi-stable motion patterns of the swarm depend on values of the coupling coefficient $a$ and the delay $\tau$,
similar to the globally coupled, homogeneous case \cite{Romero2012}; in addition, there is now a dependence on $\kappa_i$
and on the fraction of missing links in $\mc{G}$. The collective motion patterns of the swarm for different values of the
parameters $a$ and $\tau$ are described in more detail below.

\subsection{Translating state}
In the translating state, the agent locations all lie close to the swarm center of mass, and the swarm moves with
constant speed and direction. Following the calculation in \cite{Romero2012}, it can be shown that the translation
speed of the swarm center of mass $\|\dot{\bR}\|$ must satisfy
$\|\dot{\bR}\|^2 = 1-\frac{a \bar d \tau}{\langle \kappa_i \rangle}$. The first-order system in Eq.~\ref{eq:ddotR}
exhibits a pitchfork bifurcation at $a \bar d \tau = \langle \kappa_i \rangle$, where the translating state
disappears.

\subsection{Ring state}
\begin{figure}[htb]
\centering
\begin{minipage}[t]{0.32\textwidth}
\includegraphics[width=\textwidth]{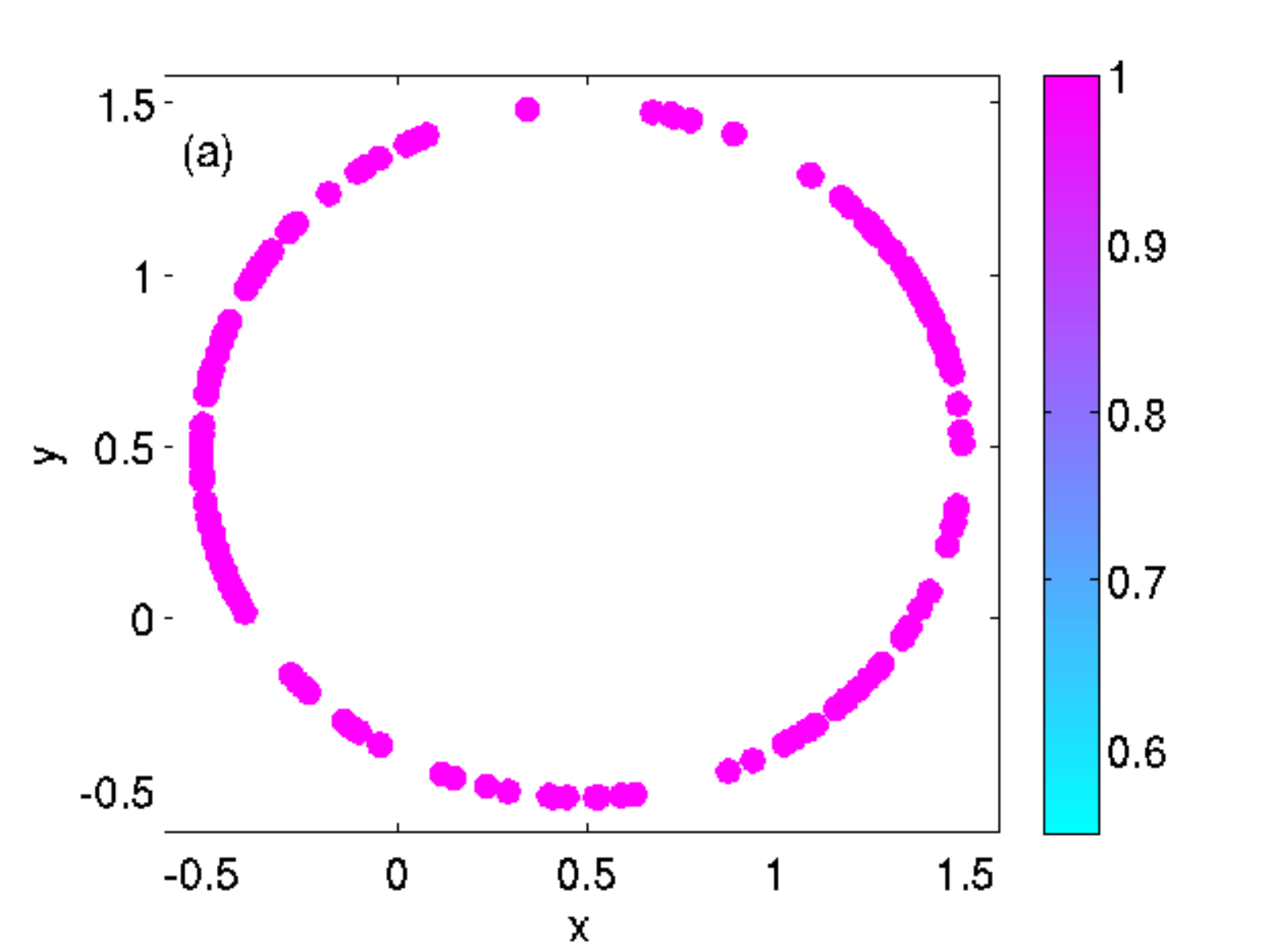}
\end{minipage}\hfill%
\begin{minipage}[t]{0.32\textwidth}
\includegraphics[width=\textwidth]{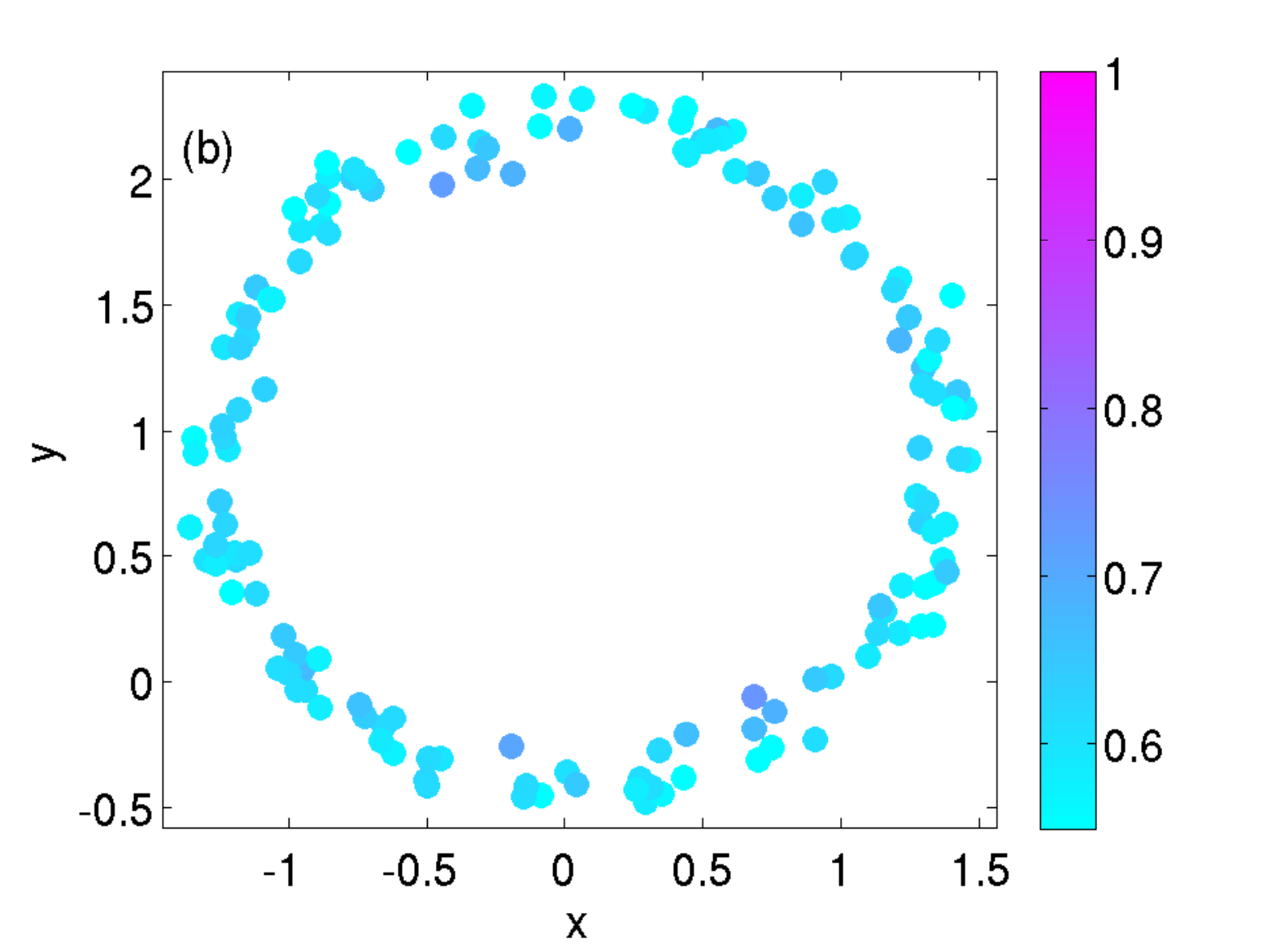}
\end{minipage}%
\caption{Snapshots of homogeneous ($\kappa_i = 1$) swarms in the ring state; (a) all-to-all communication graph,
(b) Erd\"{o}s-Renyi communication graph with $\bar d = 0.6$. Both swarms have coupling constant $a = 1$ and
communication delay $\tau = 2.5$ (color online).}
\label{fig:ringstate}
\end{figure}

For all values of $a$ and $\tau$, (\ref{eq:ddotR}) admits a stationary solution, $\mb{R}(t) = \mb{R}(0)$. In this
state, the agents form an annulus (``ring'') about the stationary center of mass; within the annulus, agents rotate in
either direction about the center. To find the radius of the annulus and angular velocity of the circling swarm agents,
we convert to polar coordinates $(\rho_i, \theta_i)$, where
$\delta \br_i = [\rho_i \cos (\theta_i),\, \rho_i \sin (\theta_i)]^T$. In the ring state, the center of mass is stationary;
without loss of generality, we set $\bR \equiv 0$; in addition, we have
$\rho_i =$ const. and $\dot \theta_i = \omega_i= $ const. Writing Eq.~\ref{eq:ddeltar} in polar
coordinates and setting the appropriate derivatives to $0$, we get the following set of equations for the motion of
individual agents in the ring state:
\begin{align*}
 \rho_i \omega_i^2 \cos (\theta_i) &= \kappa_i (1-\rho_i^2 \omega_i^2)\rho_i \omega_i \sin (\theta_i) \\
 & \qquad + \frac{\kappa_i a}{N} \sum_{j\in\mc{N}_i}\( \rho_i \cos (\theta_i) - \rho_i^\tau \cos (\theta_i)^\tau\) \\
 \rho_i \omega_i^2 \sin (\theta_i) &= \kappa_i (1-\rho_i^2 \omega_i^2)\rho_i \omega_i \cos (\theta_i) \\
 & \qquad + \frac{\kappa_i a}{N} \sum_{j\in\mc{N}_i}\( \rho_i \sin (\theta_i) - \rho_i^\tau \sin (\theta_i)^\tau\).
\end{align*}
For a communication graph with sufficiently high degree, the radius and angular velocity can be
approximated by
\begin{subequations}\label{ring_sol_ER}
\begin{align}
 \rho_i &= \sqrt{N/a \kappa_i |\mc{N}_i|} \label{eq:rho_i_limit} \\
 \omega_i &= \pm \sqrt{a \kappa_i |\mc{N}_i|/N}
\end{align}
\end{subequations}
(see Fig.~\ref{fig:ringstate} for an illustration in the case of all-to-all coupling and heterogeneous acceleration coefficients).

The stability of the ring state is determined by the eigenvalues associated with the characteristic equation associated
with the system in Eq.~\ref{eq:ddotR},
\begin{equation}
 M(\lambda; a,\tau) = \left[ \lambda^2 - \langle \kappa_i \rangle \lambda + a \bar d \(1-e^{-\lambda \tau}\) \right]^2.
\end{equation}
The ring state loses stability in a Hopf bifurcation; solving for where roots of $M$ cross the imaginary axis,
we obtain a family of Hopf bifurcation curves in the $a-\tau$ parameter space:
\begin{equation}
 \tau = \frac{1}{\sqrt{2 a \bar d - \langle \kappa_i \rangle}}\left( \arctan \frac{\langle \kappa_i \rangle\sqrt{2a \bar d - \langle \kappa_i \rangle}}{\langle \kappa_i \rangle^2-a \bar d} +
 2 m \pi\right),
\end{equation}
where $m \in \mathbb{Z}$. The rotating state, in which the agents collapse and collectively rotate about a fixed point,
is created along the curve corresponding to $m=0$.

\begin{figure}[htb]
 \centering
 \includegraphics[width=.45\textwidth]{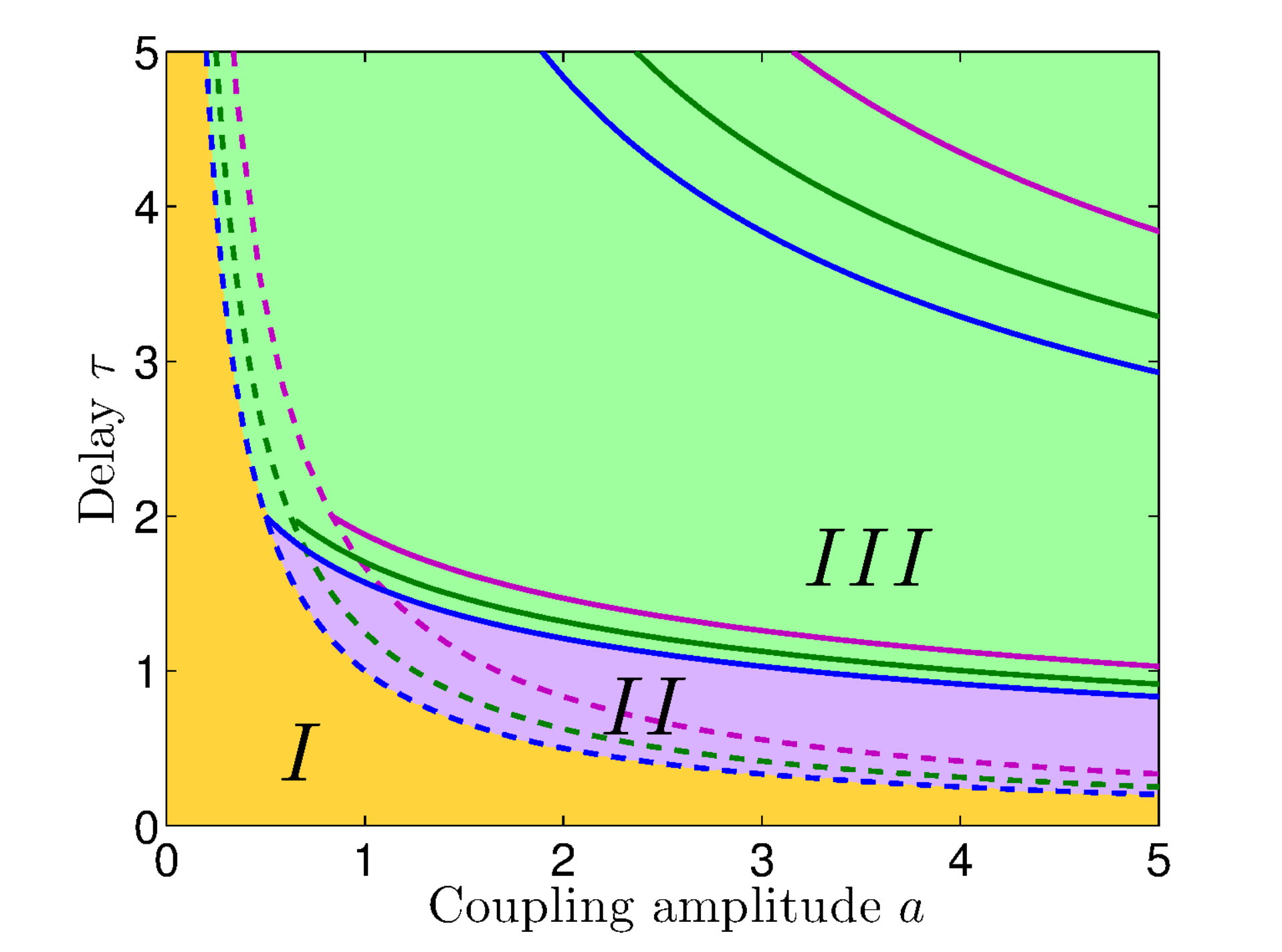}
 \caption{Bifurcation curves for $\bar{d}=1.0$ (blue), $\bar{d}=0.8$ (green), and $\bar{d}=0.6$ (violet)
 (color online). \revision{The solid lines represent families of Hopf bifurcation curves. Due to the delay term in the 
 mean-field equations of motion, there is an infinite family of Hopf bifurcations, though they do not necessarily 
 correspond to formation of new stable physical states.} The dashed lines show the 
 location of the pitchfork bifurcation where the translating state disappears. \revision{Note that as $\bar{d}$ is 
 decreased, the curves shift to the right.} In the mean field, the translating state occurs in region I; the ring state
 in region II, and the rotating state, in region III \revision{(see supplementary materials for video showing simulation 
 of the three collective motion states)}.}
 \label{fig:avtau}
\end{figure}

When $\bar d=1$, we recover the equations for the globally coupled system. The factor of $\bar d$ in which results from
breaking a fraction of the links in the global network represents a perturbation from the globally-coupled case. The
result is a shift in the bifurcation curves, as shown in Fig.~\ref{fig:avtau} for the homogeneous case
$\kappa_i = 1$. The pitchfork and Hopf bifurcation curves meet at a Bogdanov-Takens bifurcation point when
$a = \frac{\langle \kappa_i \rangle^2}{2\bar d}$, $\tau = \frac{2}{\langle \kappa_i \rangle}$.

\subsection{Rotating state}
\begin{figure}[htb]
\centering
\includegraphics[width=0.4\textwidth]{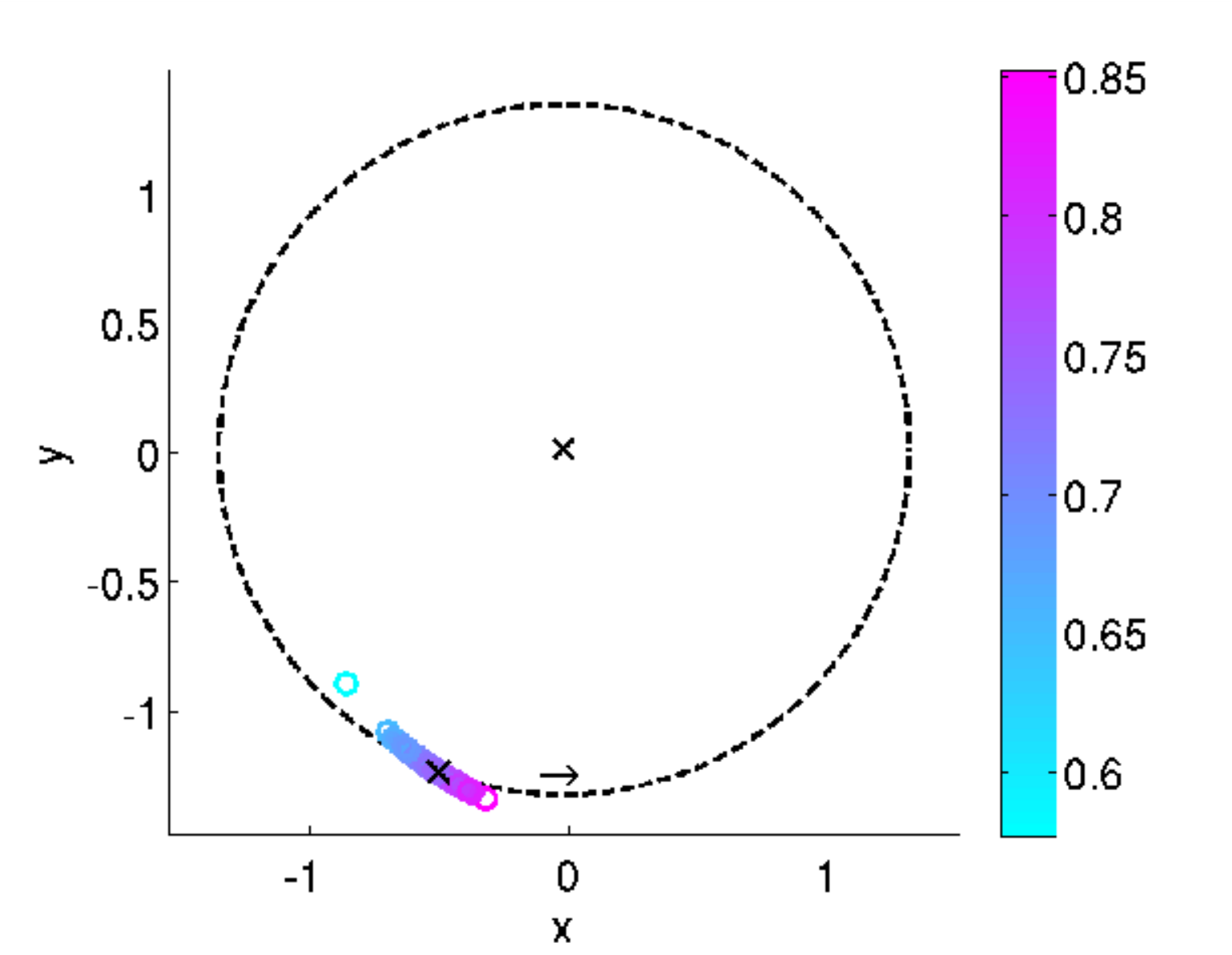}
\caption{Snapshot of simulation showing a homogeneous swarm in the rotating state \revision{at $t=3000$ (in normalized units)}.
The stationary center point is marked by `$\times$'. The colors correspond to the normalized coupling degree
$|\mc{N}_i|/(N-1)$. Here $a = 1$, $\tau = 4.5$, and $\bar d = 0.75$ (color online).}
\label{fig:rotex}
\end{figure}

\begin{figure}[htb]
\centering
\begin{minipage}[t]{0.33\textwidth}
\includegraphics[width=\textwidth]{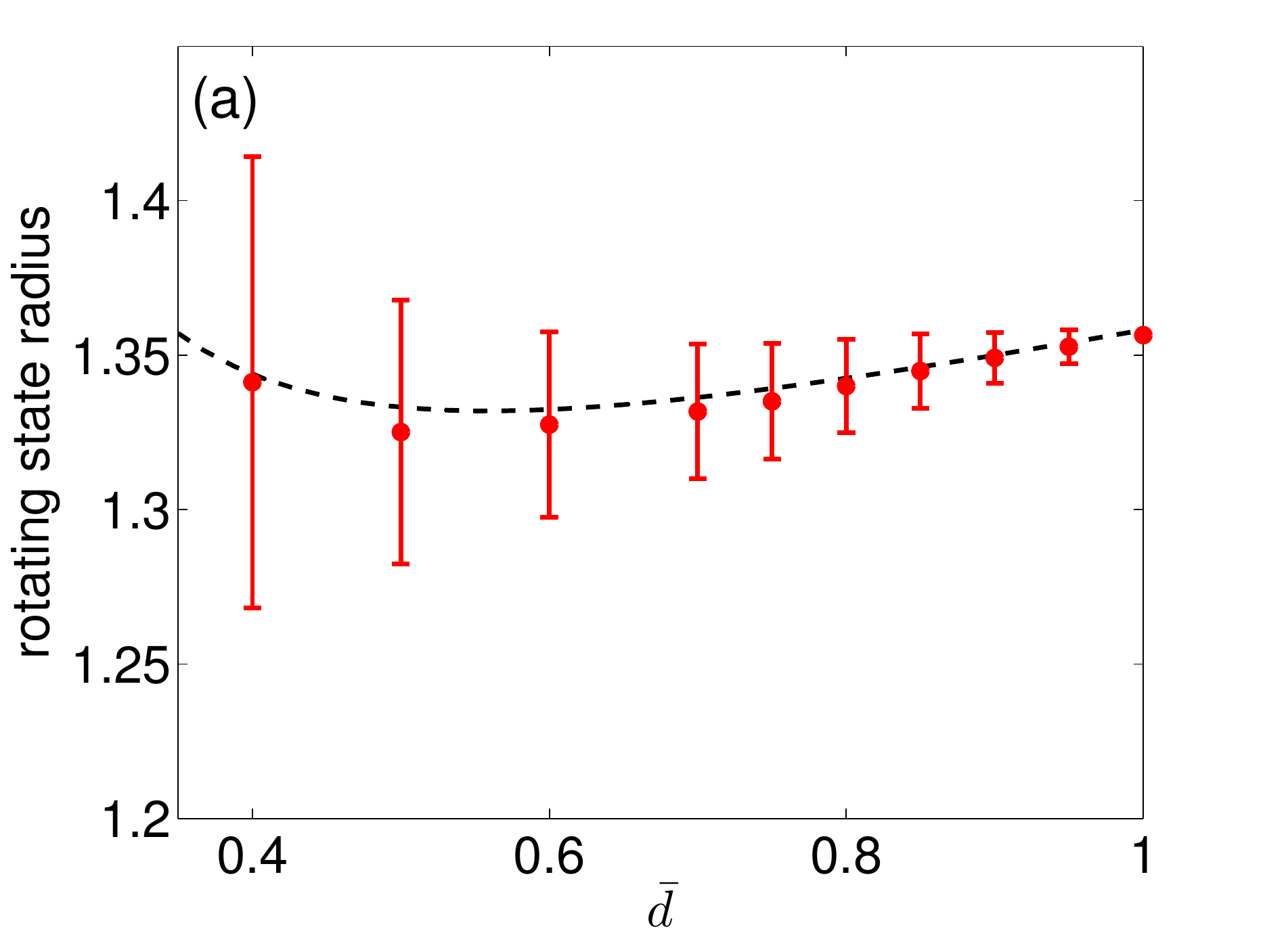}
\end{minipage}\hfill%
\begin{minipage}[t]{0.33\textwidth}
\includegraphics[width=\textwidth]{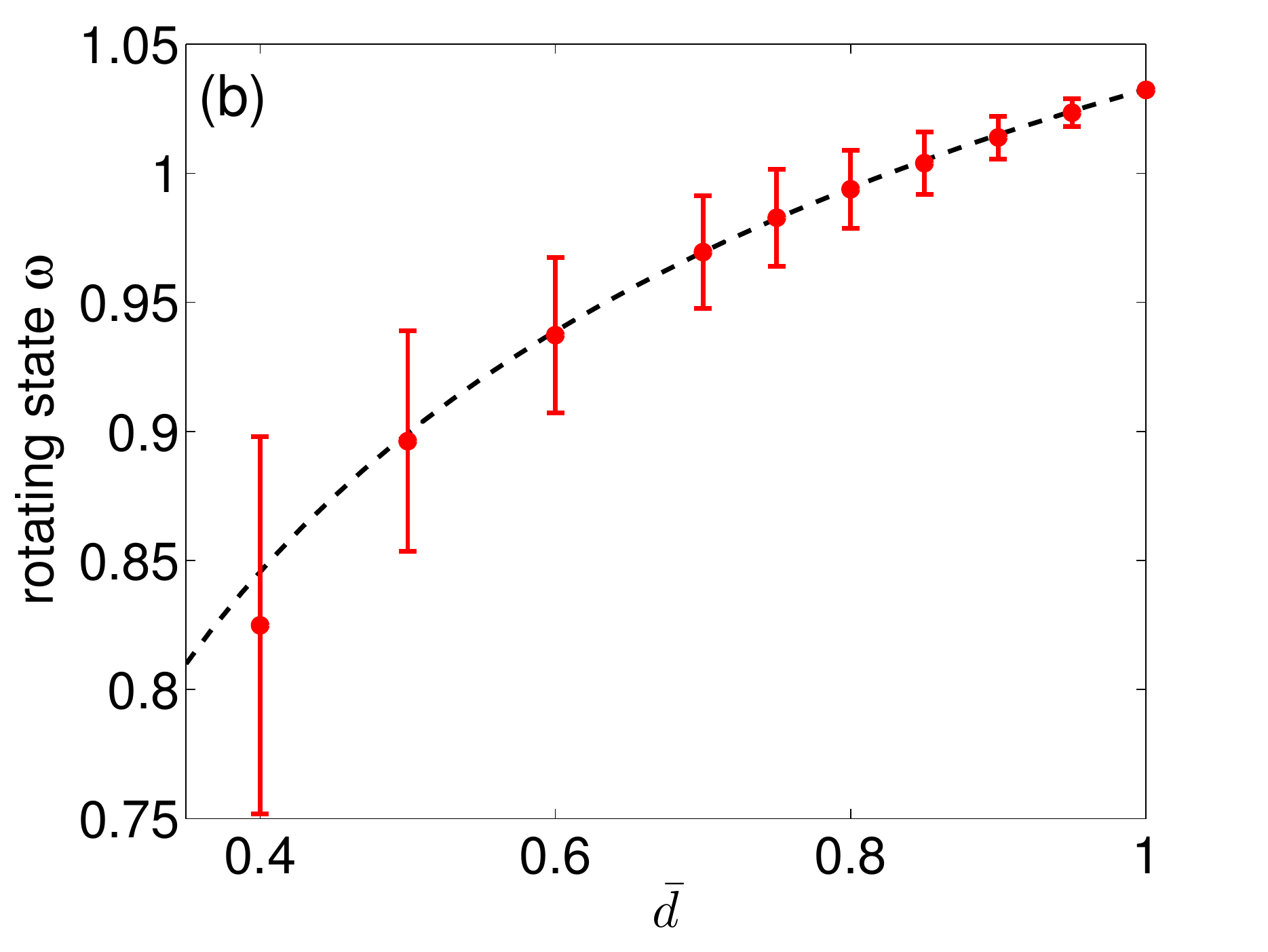}
\end{minipage}%
\caption{Comparison of the theoretical values of radius (a) and angular velocity (b) of the
agents in the rotating state with full-swarm simulations of 150 agents, with $a = 1$, $\tau = 4.5$,
and $\kappa_i = 1$, for different connection degrees. The simulation values are obtained by
averaging over all agents in the swarm\revision{, over ten simulation runs}. Error bars are shown one standard 
deviation above and below the mean values for each swarm.}
\label{fig:rotstate}
\end{figure}

In the rotating state, the agents move in a tight group about a fixed center of rotation. The radius $\rho_{\rm rot}$
and angular velocity $\omega_{\rm rot}$ of the center of mass of the swarm in the rotating state satisfy
\begin{subequations} \label{eqs:omrhoCM}
\begin{align}
 \omega_{\rm rot}^2 &= a \bar d (1-\cos (\omega_{\rm rot} \tau)) \\
 \rho_{\rm rot}  &= \frac{1}{|\omega_{\rm rot}|}\sqrt{1-\frac{a \bar d \sin (\omega_{\rm rot} \tau)}{\langle \kappa_i \rangle \omega_{\rm rot}}}
\end{align}
\end{subequations}
(see Fig.~\ref{fig:rotstate}). In the case of global coupling with homogeneous agents ($\kappa_i = 1$), all agent
positions in the rotating state coincide; however, when coupling is not global or when the agent dynamics are not
homogeneous, different agents circle the fixed point with equal angular frequency but have different radii, and have a
fixed relative phase offset from the center of mass, depending on their acceleration factor/coupling degree (see
Fig.~\ref{fig:rotex}).

We now take a closer look at the distribution of agents in the rotating state, and separately examine the effects of
having  heterogeneous agent dynamics and non-global communication. First, note that the coupling term for
agent $i$ in Eq.~\ref{eq:ddeltar} can be approximated by
\begin{align*}
\sum_{j \in \mc{N}_i}({\mb r}_i(t) - {\mb r}_j^\tau(t)) =& |\mc{N}_i| (\bm{r}_i - \bm{R}^\tau) - \sum_{j \in \mc{N}_i}\delta \bm{r}_j^\tau \\
\approx & |\mc{N}_i| (\bm{r}_i - \bm{R}^\tau),
\end{align*}
where we assume that $\delta \bm{r}_j^\tau$ are small since the system is in the rotating state (this assumption
breaks down as the degree of the communication graph $\mc{G}$ is decreased sufficiently). The equation of motion for
agent $i$ can be written as
\begin{equation*}
 \ddot{\mb r}_i = \kappa_i(1-\|\dot{\mb r}_i\|^2)\dot{\mb r}_i - \frac{a \kappa_i |\mc{N}_i|}{N}({\mb r}_i(t) - {\mb R}^\tau(t)).
\end{equation*}
We let $N \> \infty$ while keeping the ratio $|\mc{N}_i|/N$ constant. Let $d_i$ be defined as $\kappa_i |\mc{N}_i|/N$,
so that
\begin{equation}
 \ddot{\mb r}_i = \kappa_i(1-\|\dot{\mb r}_i\|^2)\dot{\mb r}_i - a d_i ({\mb r}_i(t) - {\mb R}^\tau(t)),
\end{equation}
where the motion of the center of mass $\mb{R}$ is given by (\ref{eq:ddotR}).

Let $(\rho_{\rm rot}, \theta_{\rm rot})$ and $(\rho_i,\theta_i)$ denote the polar coordinates of the swarm center of
mass and of agent $i$, respectively, relative to the center of rotation. In these coordinates, the motion of agent $i$
in the swarm in the rotating state $\rho_i =$ const. and $\dot \theta = \omega_i= \omega_{\rm rot}=$ const. is described by
\begin{subequations}
\begin{align}
\rho_i \omega_{\rm rot}^2 \cos (\theta_i) &= \kappa_i(1-\rho_i^2 \omega_{\rm rot}^2) \rho_i \omega_{\rm rot} \sin (\theta_i) \notag \\
& + a d_i (\rho_i \cos (\theta_i) - \rho_{\rm rot} \cos(\theta-\omega_{\rm rot})) \\
\rho_i \omega_{\rm rot}^2 \sin (\theta_i) &= \kappa_i(1-\rho_i^2 \omega_{\rm rot}^2) \rho_i \omega_{\rm rot} \cos (\theta_i) \notag \\
&  + a d_i (\rho_i \sin (\theta_i) - \rho_{\rm rot} \sin(\theta-\omega_{\rm rot})),
\end{align}
\end{subequations}
where $\rho_{\rm rot}$ and $\omega_{\rm rot}$ can be computed from Eq.~\ref{eqs:omrhoCM}. Let
$\Delta \theta_i = \theta_i - \theta_{\rm rot}$ be the phase offset between agent $i$ and the swarm center of mass. It
can be shown that $\rho_i$ and $\Delta \theta_i$ must satisfy
\begin{subequations}
\begin{align}
\rho_i &= \frac{\cos(\omega_{\rm rot} \tau + \Delta \theta_i)}{1-\frac{\bar d}{d_i}(1-\cos (\omega_{\rm rot} \tau))} \rho_{\rm rot} \\
\(1-\rho_i^2 \omega_{\rm rot}^2\) \rho_i &= \frac{a d_i \rho_{\rm rot}}{\kappa_i \omega_{\rm rot}} \sin(\omega_{\rm rot} \tau + \Delta \theta_i)
\end{align}
\end{subequations}
This set of coupled nonlinear equations can be solved numerically for $\rho_i$ and $\Delta \theta_i$ for different
values of $a$, $\tau$, $\kappa_i$ and $d_i$. We consider two cases separately. First, we assume that the swarm agents
are homogeneous, with $\kappa_i = 1$ for all $i$; in this case, $d_i = |\mc{N}_i|/N$ is the normalized degree of agent
$i$ in the communication graph $\mc{G}$, and $\bar d$ is the mean degree. Solution curves for $\rho_i$ and
$\Delta \theta_i$ for different values of $d_i$ are plotted in Appendix \ref{app:myplots}. Second, we assume that the
communication network $\mc{G}$ is all-to-all, so that $\lim_{M\>\infty}|\mc{N}_i|/N = 1$, but that agents in the swarm
have heterogeneous dynamics. In this case, $d_i = \kappa_i$ and $\bar d$ is the mean acceleration factor. The solution
curves for this case are shown in Appendix \ref{app:myplots}.

\begin{figure*}[htb]
\centering
\begin{minipage}[t]{0.35\textwidth}
\includegraphics[width=\textwidth]{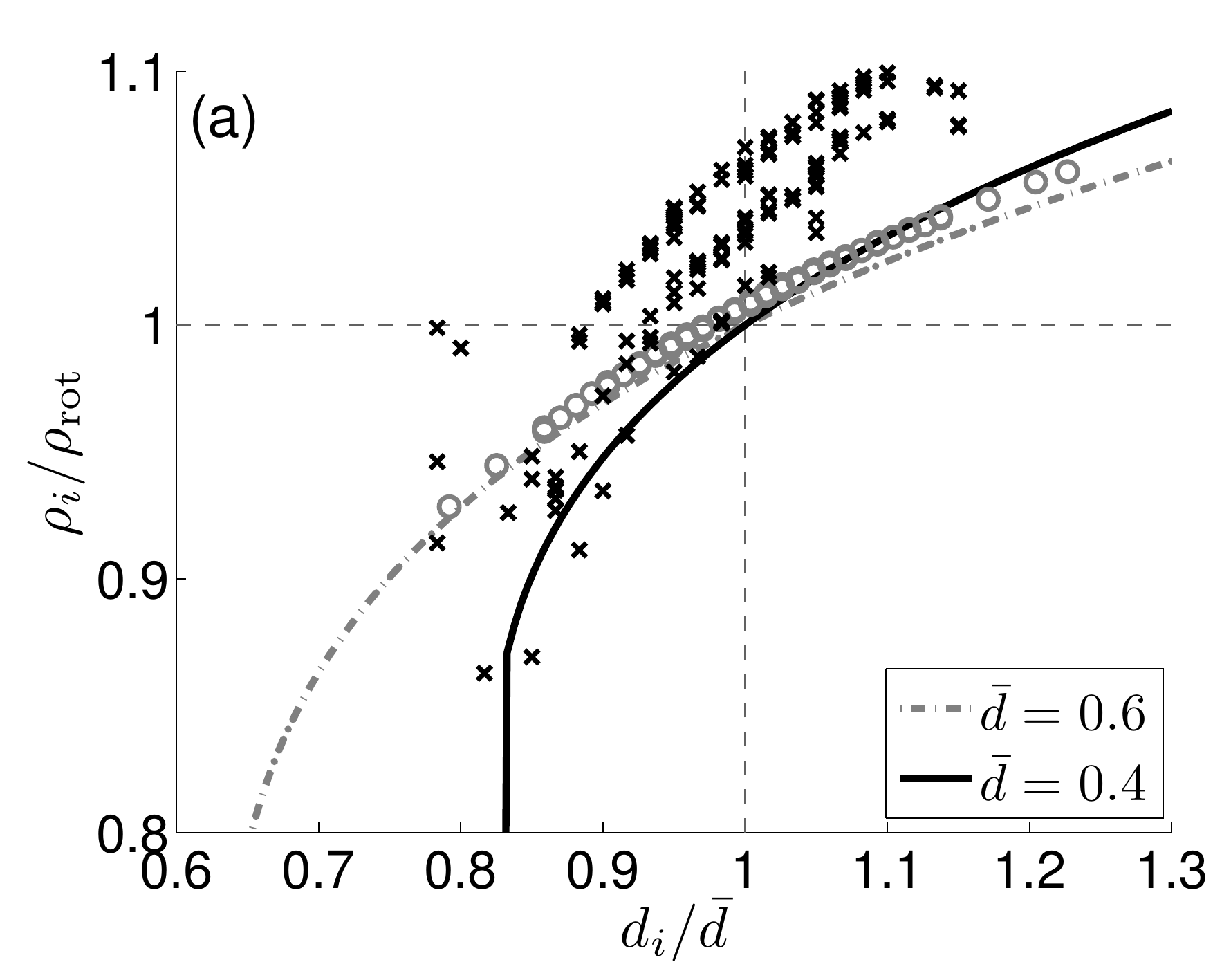}
\end{minipage}\hspace{5mm}%
\begin{minipage}[t]{0.35\textwidth}
\includegraphics[width=\textwidth]{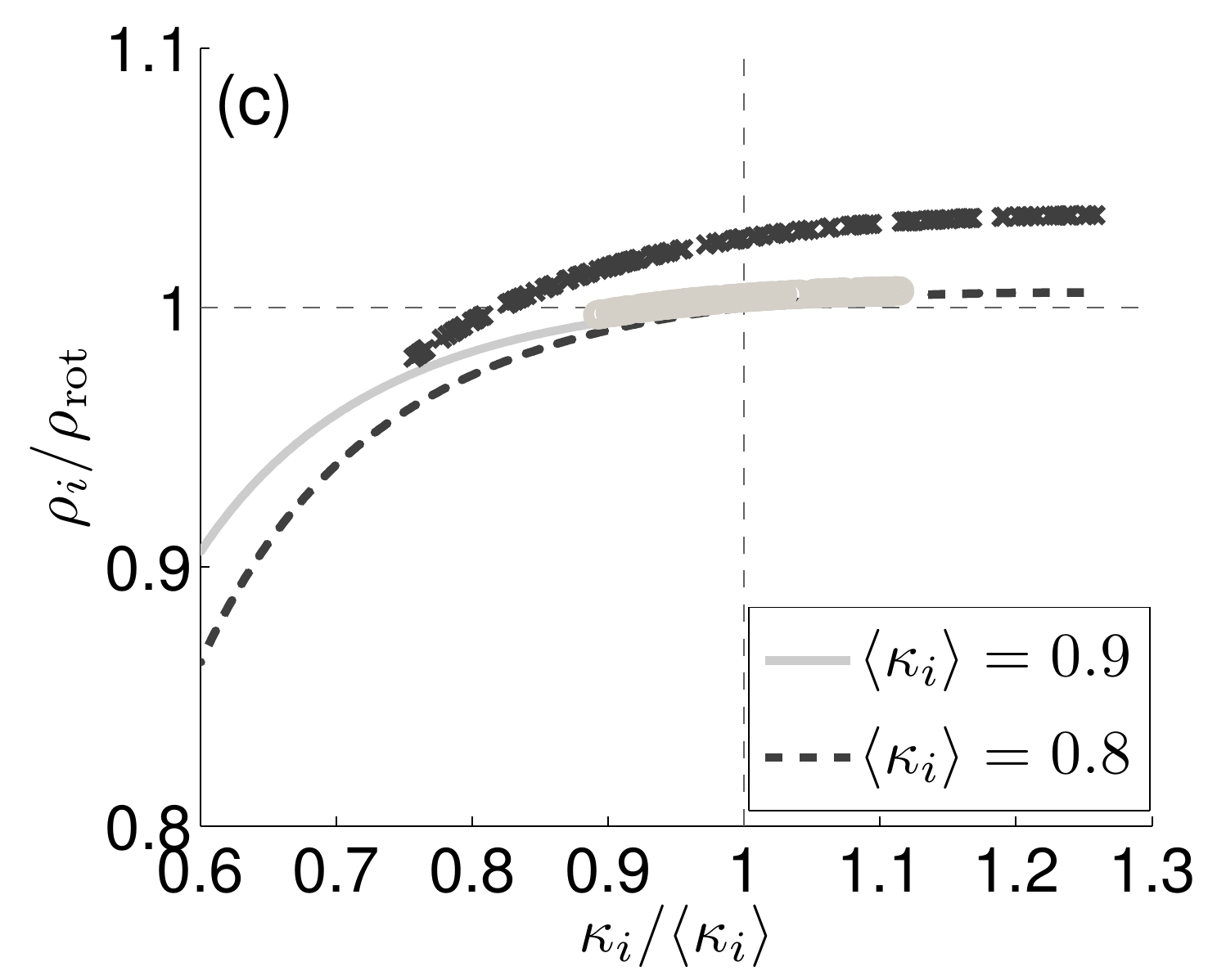}
\end{minipage}
\\
\vspace{5mm}
\begin{minipage}[t]{0.35\textwidth}
\includegraphics[width=\textwidth]{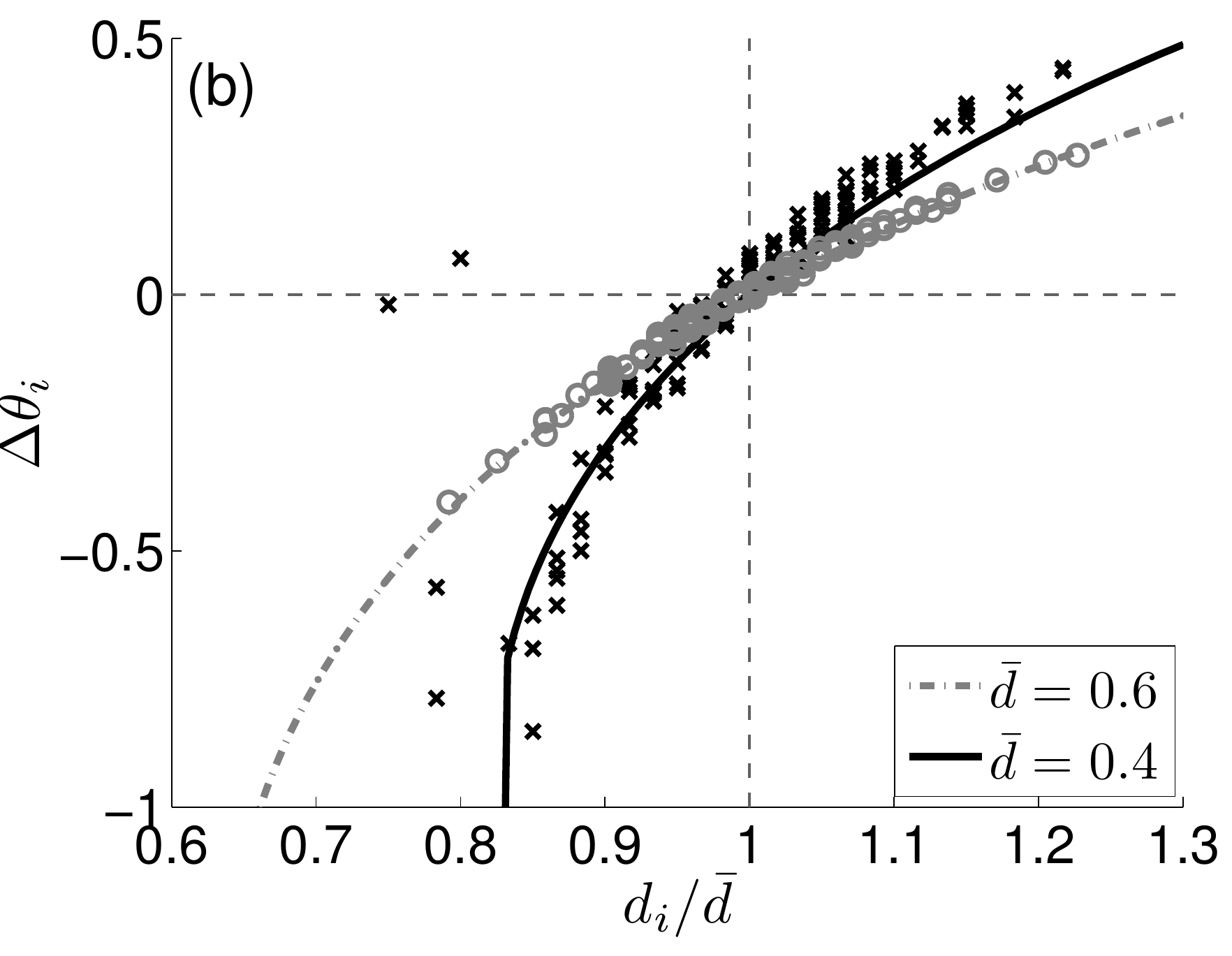}
\end{minipage}\hspace{5mm}%
\begin{minipage}[t]{0.35\textwidth}
\includegraphics[width=\textwidth]{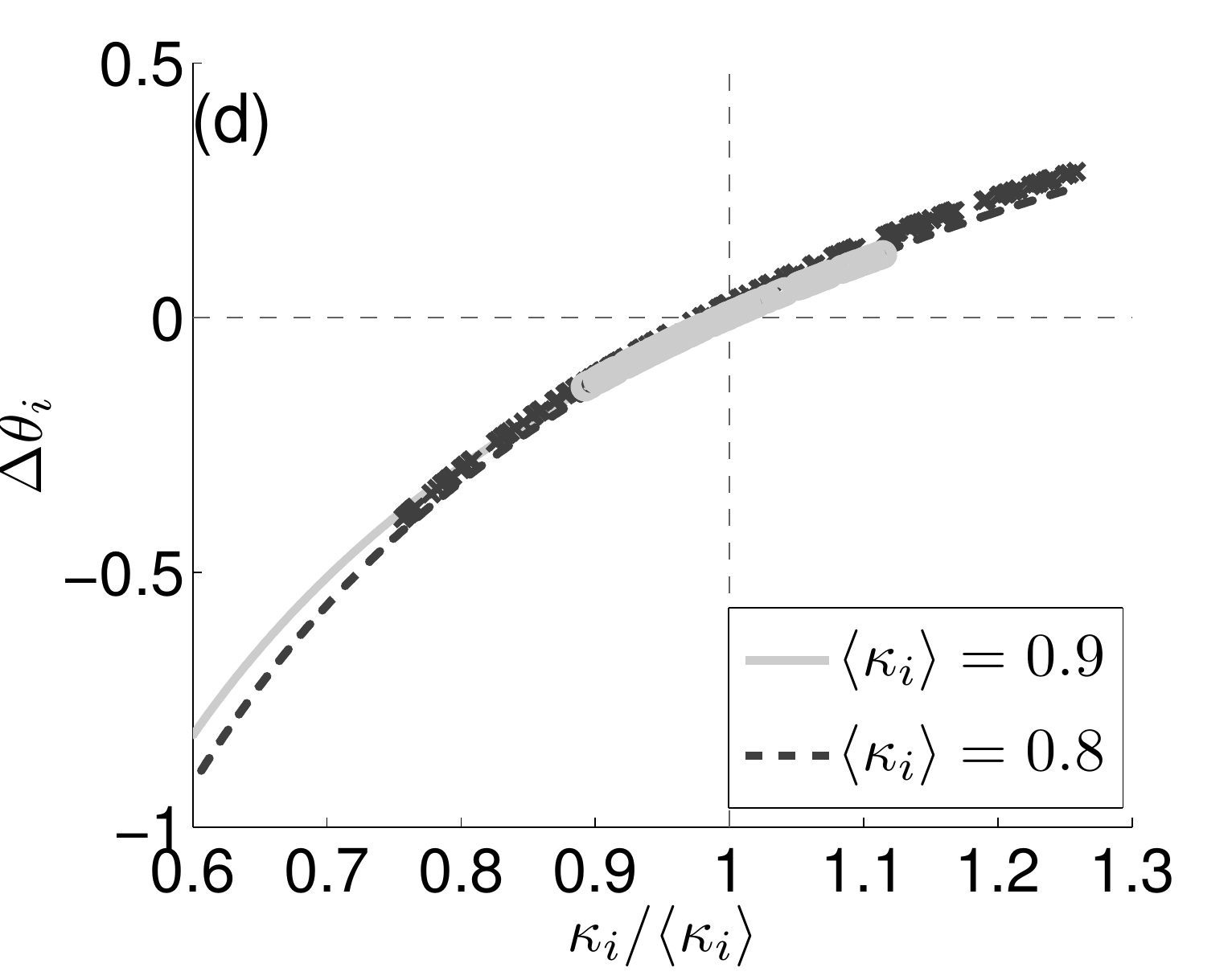}
\end{minipage}%
\caption{Comparison of simulation and theoretical values for radius (a,c) and phase difference from the center of mass
(b,d) for agent $i$. Results are shown for two sets of swarms, the first with $\bar d = 0.4$ and $0.6$ and $\kappa_i=1$
for all $i$ (a-b) and the second with all-to-all communication and $\langle \kappa_i \rangle = 0.8$ and $0.9$
(c-d). For both swarms, $a = 1$, $\tau = 4.5$, and the number of agents is $150$ (color online).}
\label{fig:twopopthm}
\end{figure*}

A direct comparison with simulation results is shown in Fig.~\ref{fig:rotstate} and Fig.~\ref{fig:twopopthm}. The slight
discrepancy in the rotating state radius in Fig.~\ref{fig:rotstate} and in $\rho_i/\rho_{\rm rot}$ in
Fig.~\ref{fig:twopopthm} (a,c) is understood as follows. Eq. (\ref{eqs:omrhoCM}) for the radius of the center of mass
assumes that agent positions deviate only slightly from the center of mass. However, as the mean coupling coefficient
decreases, or as the acceleration factors of agents in the swarm become increasingly heterogeneous, the agents become
spread out over an extended arc  (as seen in Fig.~\ref{fig:rotex}) and that assumption
becomes invalid. In this `arc'  configuration the center of mass of the swarm moves closer towards the center of
rotation than theory predicts. The analogue to a system with perturbed coupling coefficient breaks down here; for a
globally-connected swarm with decreasing coupling coefficient $a$, the rotating state disappears when the system
crosses the curve $\bar{a} \tau^2 = 2$, where the rotating state radius diverges. The swarm then transitions to a
translating state. It is, however, remarkable, that the mean-field analysis captures so much of the overall swarm
behavior even as the coupling degree is significantly decreased.

\section{Experimental realization}

\begin{figure}[htb]
 \centering
 \vspace{5mm}
 \includegraphics[width=0.4\textwidth]{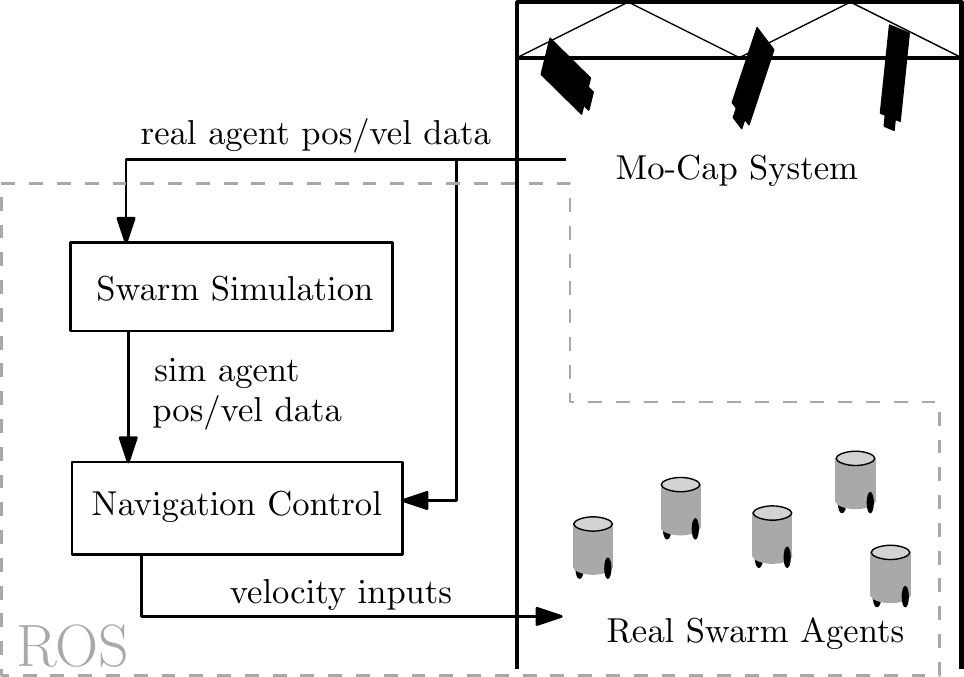}
 \caption{Experimental setup with virtual swarm. The real robots operate in a lab testbed. Positions are measured using
 an overhead motion capture system (Optitrack). The positions of the real and simulated agents are passed to the
 virtual swarm simulator, which models the response of the virtual swarm agents to the current swarm configuration; and
 to the controller, which computes the real robot response and passes target velocities to the real swarm agents. Delay
 is blown into the system artificially to simulate delays in real communication systems. Aside from Optitrack, all
 parts of the experimental system communicate with each other through ROS (robot operating system).}
 \label{fig:expsetup}
\end{figure}

We validate our theoretical results for the homogeneous swarm with all-to-all communication, using a mixed-reality
setup in which a small number of physical robots interacts with a larger virtual swarm (see Fig. \ref{fig:expsetup}).
We adopt the mixed-reality paradigm so that we can observe motion pattern formation for large swarms in a limited lab
space, without having to resolve significant logistical issues including setting up communication between
large numbers of individual agents.

\begin{figure}[htb]
 \centering
 \begin{minipage}[t]{0.32\textwidth}
\includegraphics[width=\textwidth]{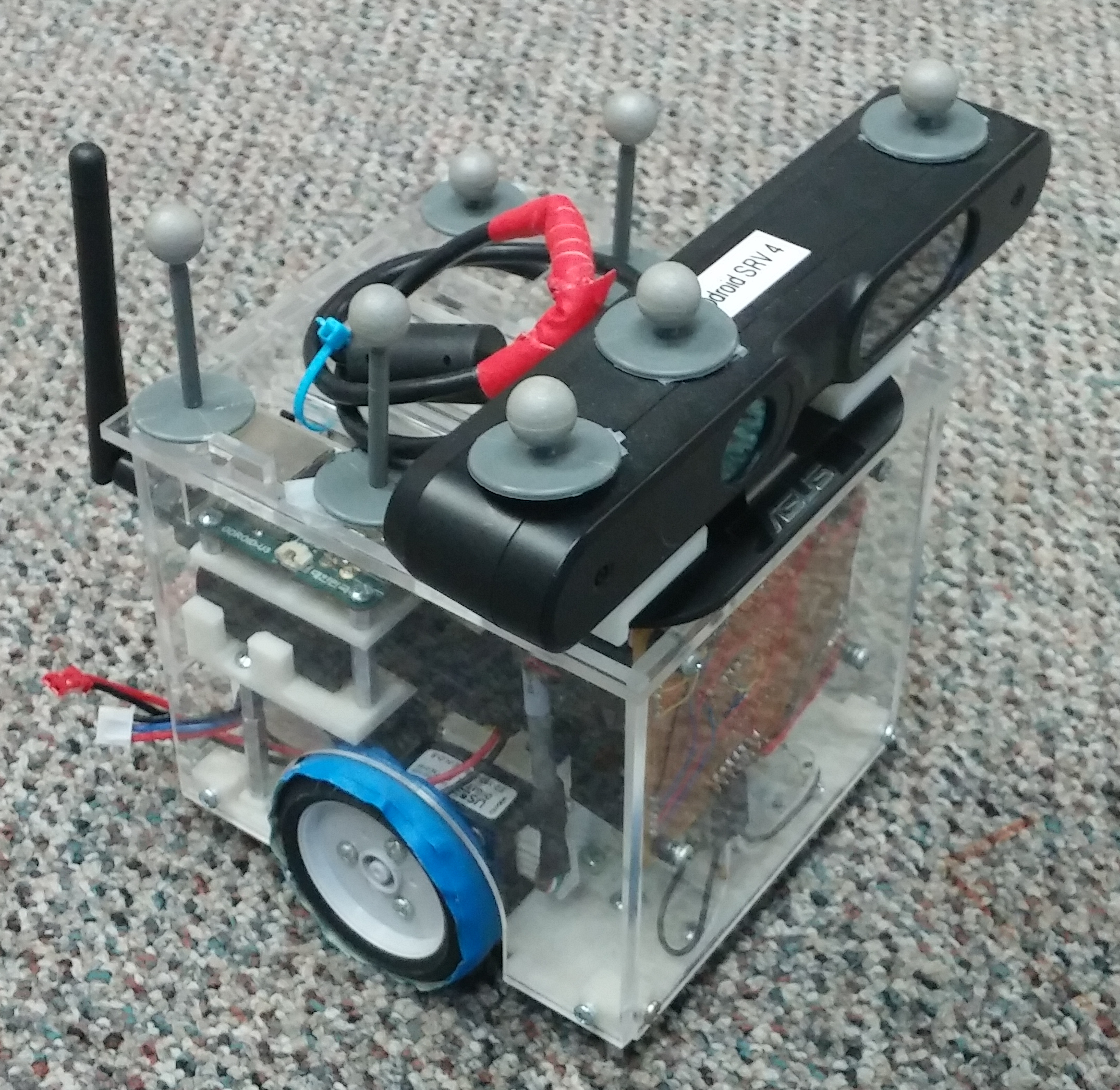}
\end{minipage}
\\
\vspace{1mm}
\begin{minipage}[t]{0.32\textwidth}
\includegraphics[width=\textwidth]{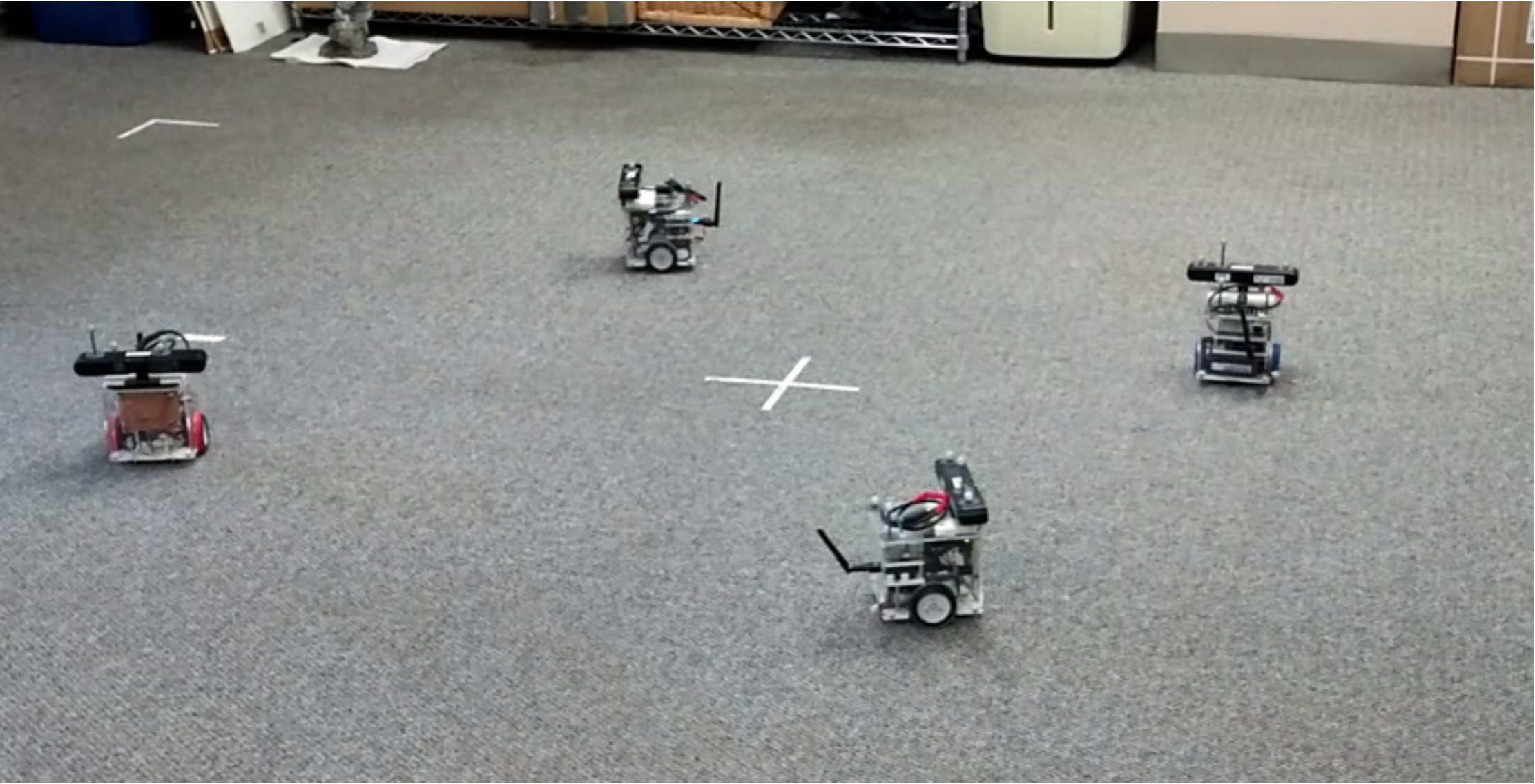}
\end{minipage}%
 \caption{Experimental platform, consisting of 2-wheeled robot cars. The top speed of the vehicles is approximately $12$ cm/s.
 Each vehicle is outfitted with reflective balls (shown) for overhead motion capture.}
 \label{fig:odroids}
\end{figure}

\begin{figure}[htb]
\centering
 \begin{minipage}[t]{0.34\textwidth}
\includegraphics[width=\textwidth]{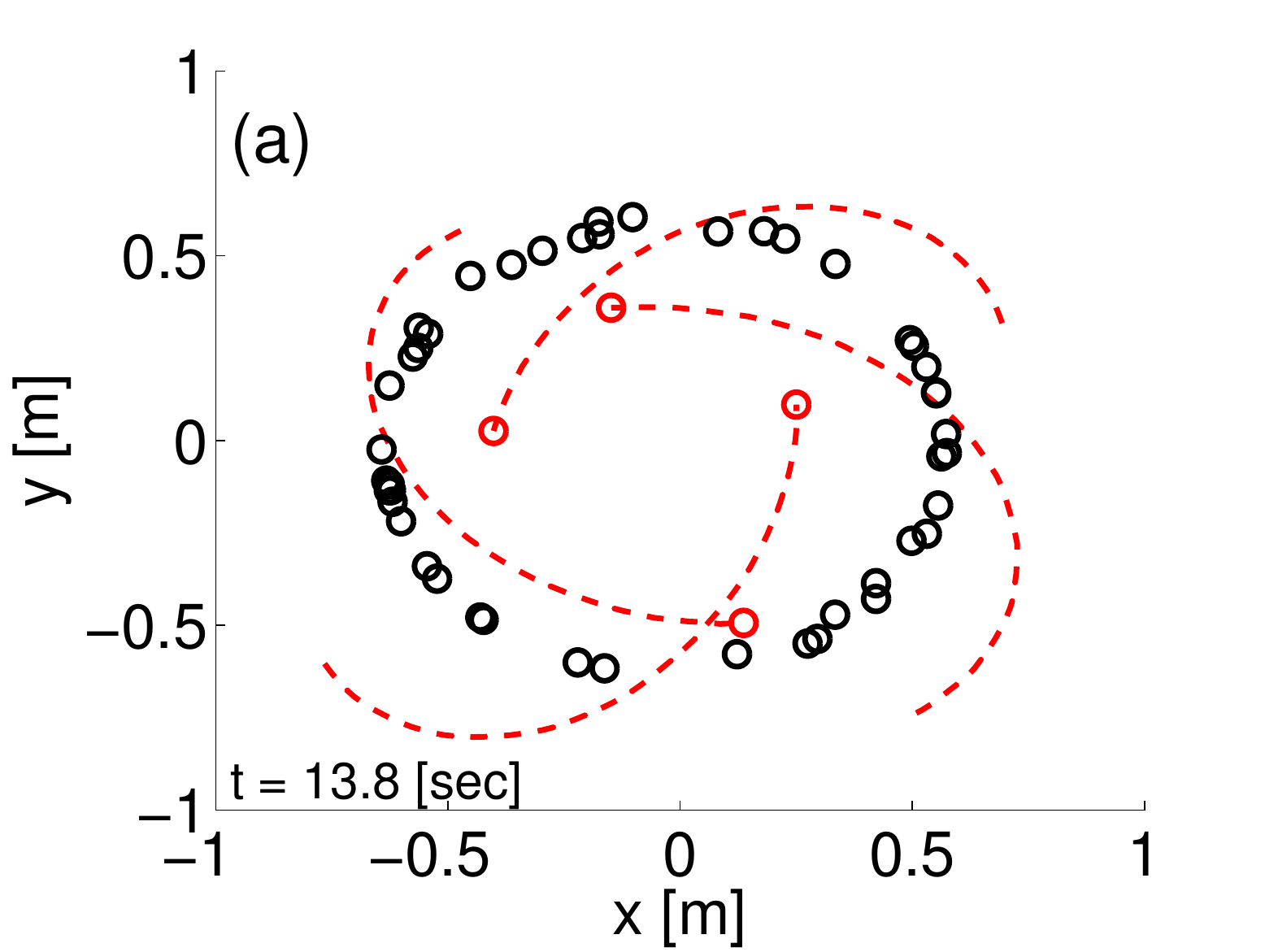}
\end{minipage}%
\\
\begin{minipage}[t]{0.34\textwidth}
\includegraphics[width=\textwidth]{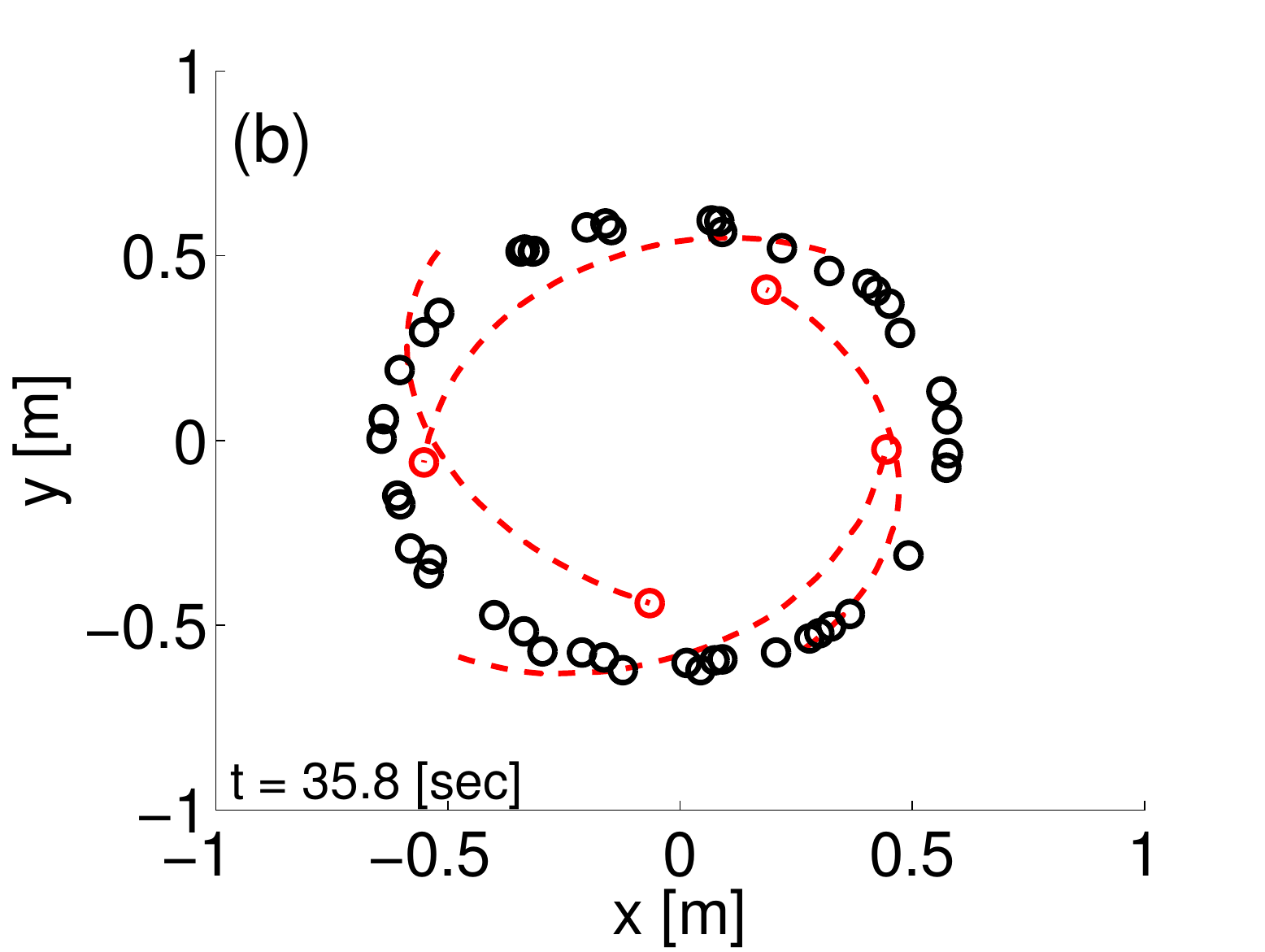}
\end{minipage}%
\\
\begin{minipage}[t]{0.34333\textwidth}
\includegraphics[width=\textwidth]{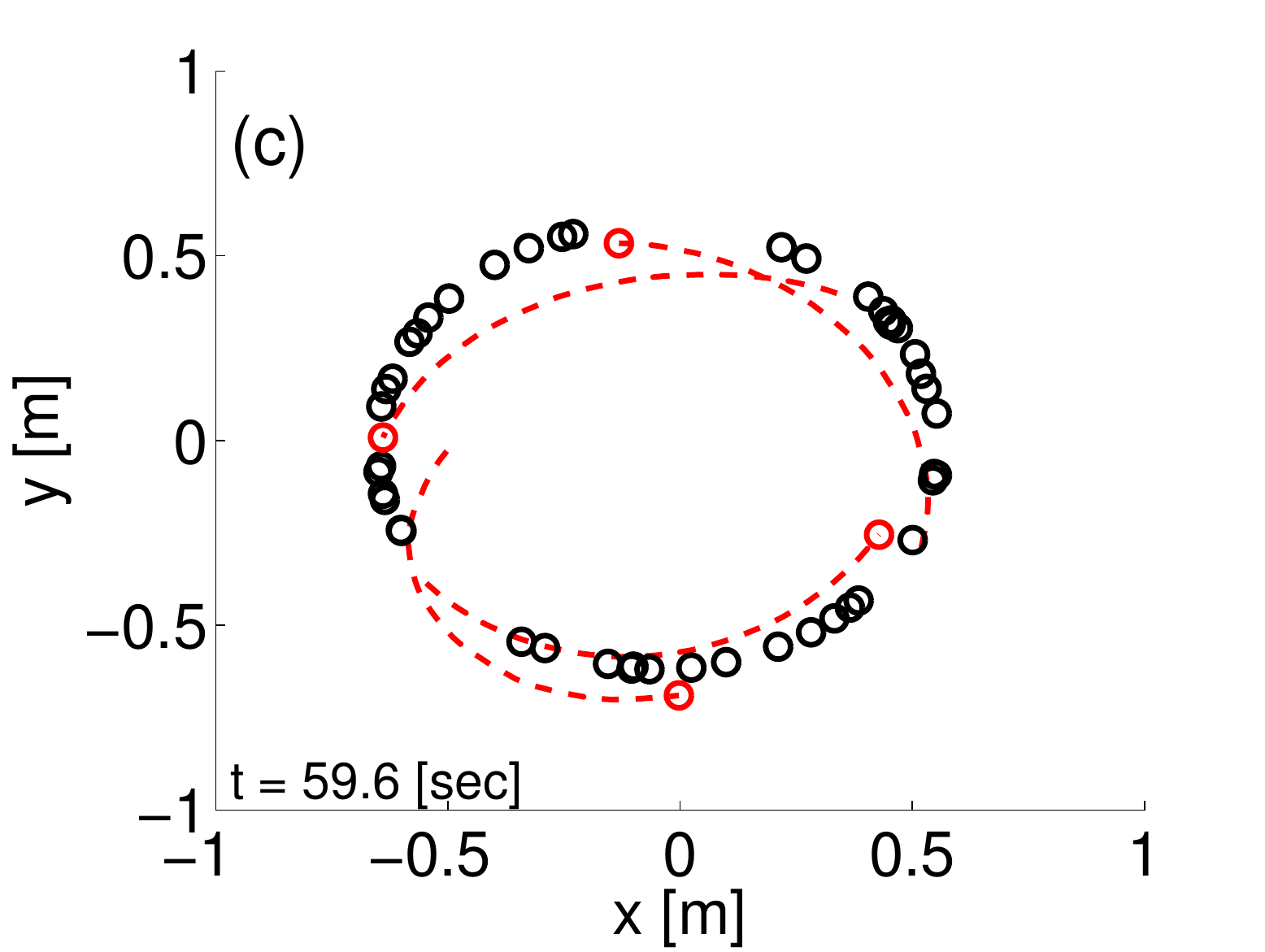}
\end{minipage}%
\caption{Snapshots of test with four real and 46 virtual agents converging to ring state. The virtual agent positions
are shown in black; the real ones in red (color online). The dashed lines represent the ten-second histories of the real robot
positions. The predicted ring radius for this experiment (coupling parameter $\alpha = 0.04\,s^{-2}$, delay $\tau = 2.5\,s$,
goal velocity $v_0 = 0.12\,m/s$, and $\beta = 20.0\,s/m^2$) was $0.6\,m$; the measured radius was $6.01\,m$.}
\label{fig:exp1}
\end{figure}

We evaluate the theory using an indoor laboratory experimental testbed  consisting of four autonomous ground vehicles
(AGVs). The AGVs are differential drive surface vehicles equipped with an Odroid U3 computer, an Xtion RGB-D sensor,
odometry, and 802.11 wireless capabilities (see Figure~\ref{fig:odroids}). Localization for each robot is provided by
an external motion capture system. By artificially adding delay in the recorded
robot positions, we simulate the effect of slow communication over a network
in the field. The (delayed) positions are passed to a simulator, which uses them (along with delayed
positions of virtual robotic agents) to update the positions of virtual agents in the swarm. In addition, the (delayed)
real and virtual robot positions are used to generate desired velocity values for the real swarm agents. The target
velocity data is passed to the real robots, and an internal PID control is applied in order to reach the target
velocities. To avoid collisions, we add repulsion between the real swarm agents. Experimental results, with 4 real
agents in ring state, are shown in Fig. \ref{fig:exp1}.

To connect the theory with experimental realization, it is necessary to dimensionalize the swarming equations, so as to
allow for actuation limits of the experimental platform. We therefore introduce a target velocity $v_0$ with units of
[length]/[time]; a dimensional coupling parameter $\alpha$ with units of 1/[time]\textsuperscript{2}; and a dimensional
factor $\beta$ with units of [time]/[length]\textsuperscript{2}. The equation governing the motion of agent $i$ can now
be expressed as:
\begin{equation}
 \ddot{\mb r}_i = \kappa_i \beta(v_0^2-\norm{\dot{\mb r}_i}^2)\dot{\mb r}_i - \kappa_i \frac{\alpha}{N} \sum_{j \in \mc{N}_i} (\bx_i - \bx_j^\tau) + \mb{F}^{\rm rep}_i,
\end{equation}
where $\mb{F}^{\rm rep}_i$ is the repulsion force on agent $i$, which acts only between real agents, and is turned on
when two agents come within a threshold distance of each other. The non-dimensional equations can be recovered by
rescaling as follows:
\begin{subequations}
\begin{align}
t' =& \beta v_0^2 \cdot t,\\
\mb{r}'_i =& \beta v_0 \cdot \mb{r}_i,
\end{align}
\end{subequations}
and
\begin{align}
a = \frac{\alpha}{\beta^2 v_0^4}.
\end{align}

We ran our experiment with parameter values $\alpha = 0.04\,s^{-2}$, $\beta = 20.0\,s/m^2$, $v_0 = 0.12\,m/s$, and time
delay $\tau = 2.5\,s$. Repulsion between real agents was switched on when they came within $0.15\,m$ of each other. For
these parameter values, we predict a ring with radius equal to $0.6\,m$; the measured radius of the ring state in this
case was $0.601\,m$. Time Snapshots of the agents converging to the ring state are shown in Fig. \ref{fig:exp1}.

Our experiment demonstrates that pattern formation can be achieved with a swarm of 50 agents (4 real and 46 virtual
agents). However, we would like to perform swarming in a truly physical environment, with all agents corresponding to
true physical robots. As a preliminary step, we have conducted a series of numerical simulations aimed at determining
the effect of finite swarm size on the pattern-formation behaviors that we analyzed in the thermodynamic limit
($N \> \infty$). See Appendix \ref{app:finiteN} for details.

Further experimental exploration of the full bifurcation structure is ongoing, and will be the described in an upcoming
paper.

\section{Conclusion}

In this paper we have analyzed the collective motion patterns of a swarm with Erd{\"o}s-Renyi communication network
structure and heterogeneous agent dynamics, using a mean-field approach from  statistical physics, with the assumption
that the number of agents goes to infinity. We derived  bifurcation diagrams demarcating regions of different
collective motions, for different values of  mean degree in the communication network. We showed that behaviors
described in \cite{Romero2012} for the globally-coupled swarm, namely translation, ring state, and rotation, persist
under heterogeneity in agent dynamics and as communication links are broken, even though the bifurcation curves are
shifted as coupling degree of the network decreases far from the all-to-all situation.

We derive expressions for the speed of the swarm in the translating state as a function of time delay and coupling
coefficient; for the mean radius and angular velocity of agents in the ring state; and for the angular velocity, and
individual radii and phase offsets for individual agents in the rotating state. We have verified these calculations
with simulations of the full-swarm dynamics and presented preliminary experimental results. It is remarkable that
our model reduction, which starts with $N$ second-order delay-differential equations and yields one equation of the
same type, is able to quantitatively capture so many aspects of the full swarm dynamics, even as the coupling degree of
agents within the swarm is significantly decreased.

In the case that many agents are coordinating together, limited communication bandwidth makes all-to-all communication
infeasible, and may lead to significant communication delays. By dropping the requirement for all-to-all communication
used in our previous work, the current paper brings us one step closer to understanding the physics of naturally-occurring
swarming systems, as well as a possible implementation of swarming control
algorithms for very large aggregates of agents. Understanding the natural emerging dynamics of the system in these
circumstances allows us to exploit them when designing controls for swarming applications.

The current work opens up interesting new areas for future study. As a next step, we plan to examine the dynamics
of swarm formation with pulsed communications, and in the presence of external disturbances (e.g.. ambient flow for
swarms of autonomous underwater agents in dynamic flow environments, such as the ocean). We also plan to conduct more
extensive experimental verification of our results. We will test how our results scale with the number of agents in the
network, and apply parametric control for dynamic pattern-switching.

\section*{Acknowledgments}

This research is funded by the Office of Naval
Research (ONR).  KS and IBS  are supported by ONR
Contract No. N0001412WX20083 and NRL Base Funding Contract No.
N0001414WX00023. DM and MAH are supported by ONR Contracts No.
N000141211019 and No. N000141310731.
This research was performed while KS and CRH held a National Research Council Research Associateship Award at the U.S. Naval
Research Laboratory.    LMR is a post-doctoral fellow at Johns Hopkins
University supported by the National Institutes of Health.

%


\begin{appendix}

\section{Rotating state radius and phase offset}
\label{app:myplots}
\begin{figure*}[htb]
 \begin{minipage}[t]{0.32\textwidth}
\includegraphics[width=\textwidth]{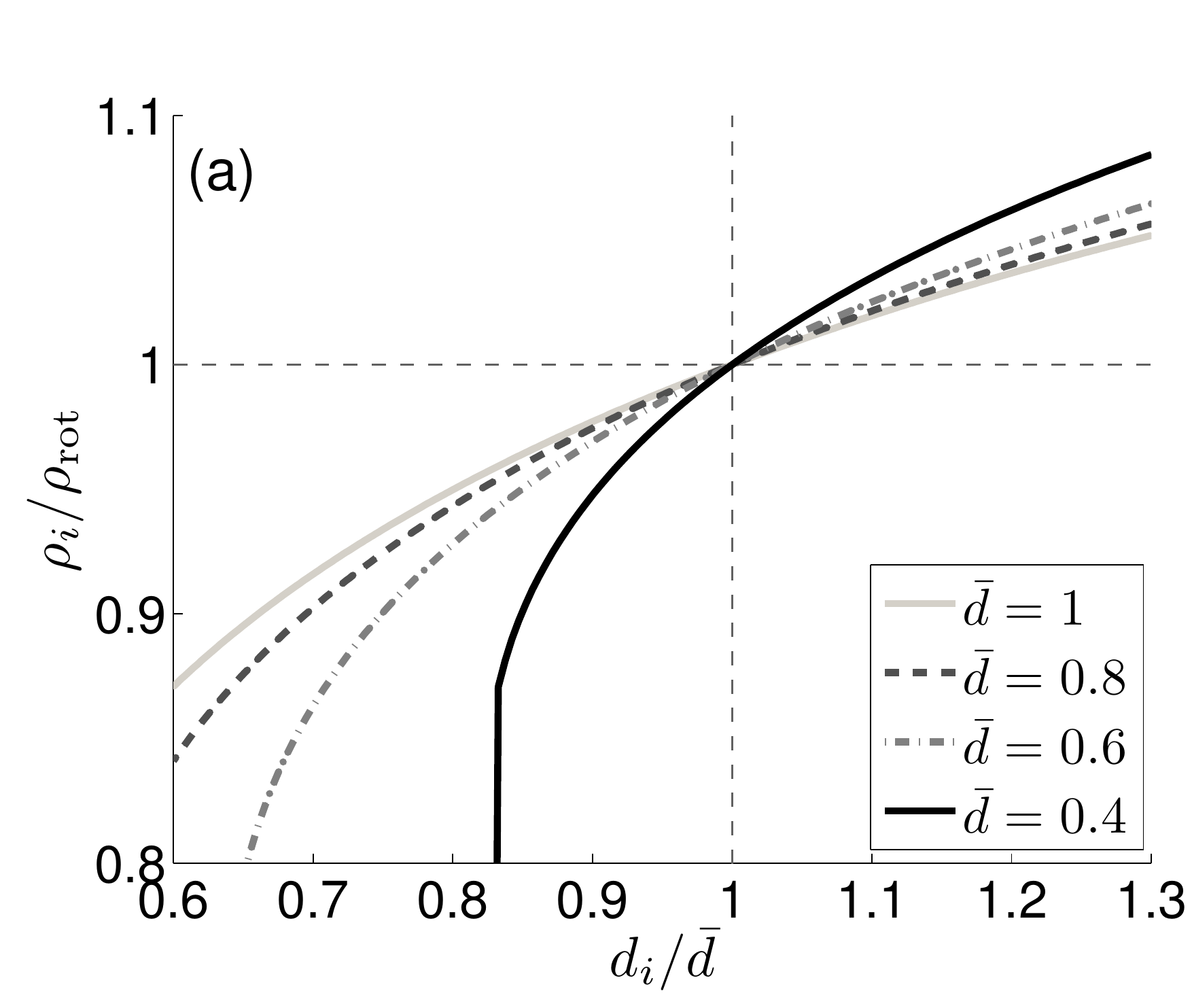}
\end{minipage} \hspace{5mm} %
\begin{minipage}[t]{0.32\textwidth}
\includegraphics[width=\textwidth]{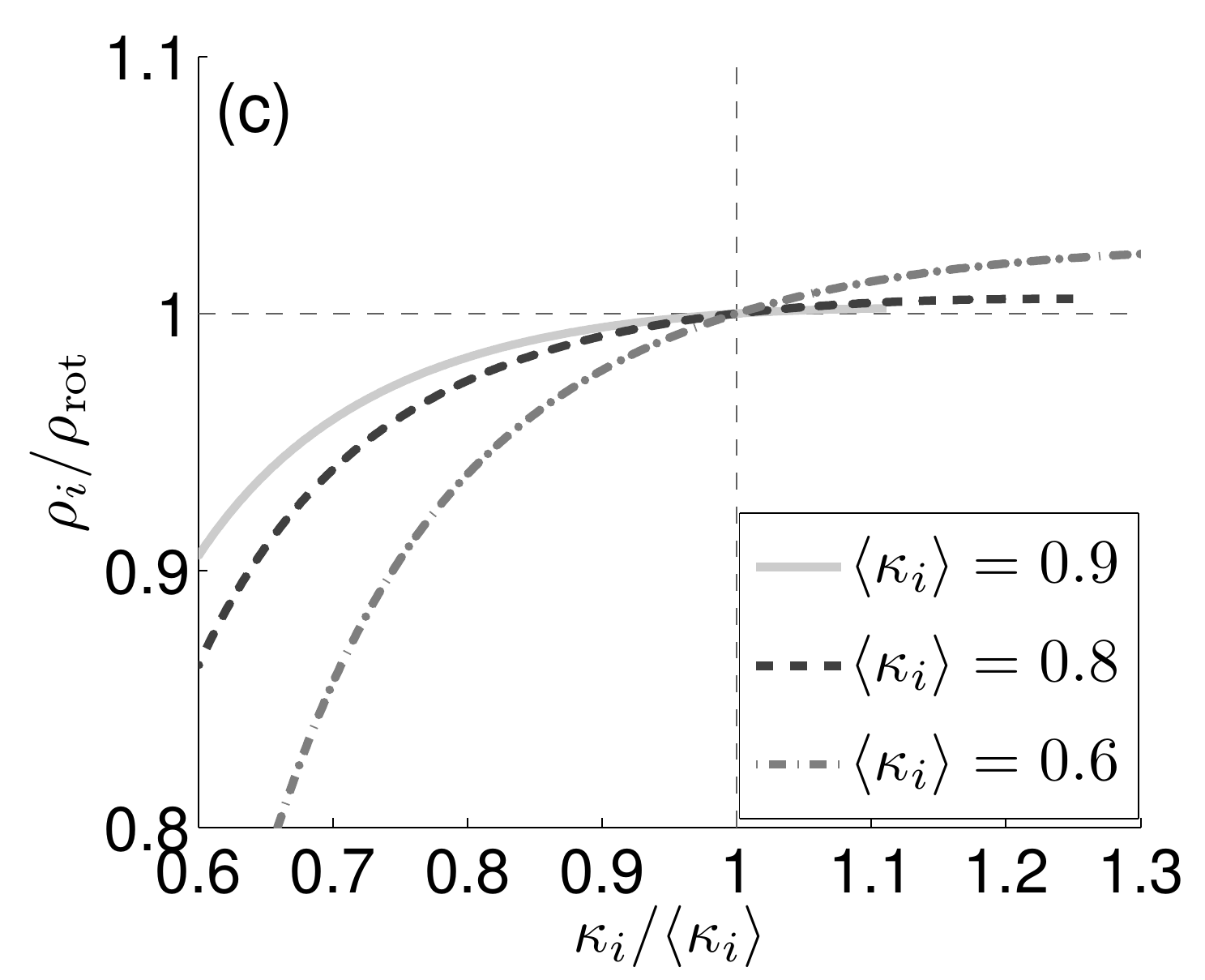}
\end{minipage}%
\\
\vspace{5mm}
\begin{minipage}[t]{0.32\textwidth}
\includegraphics[width=\textwidth]{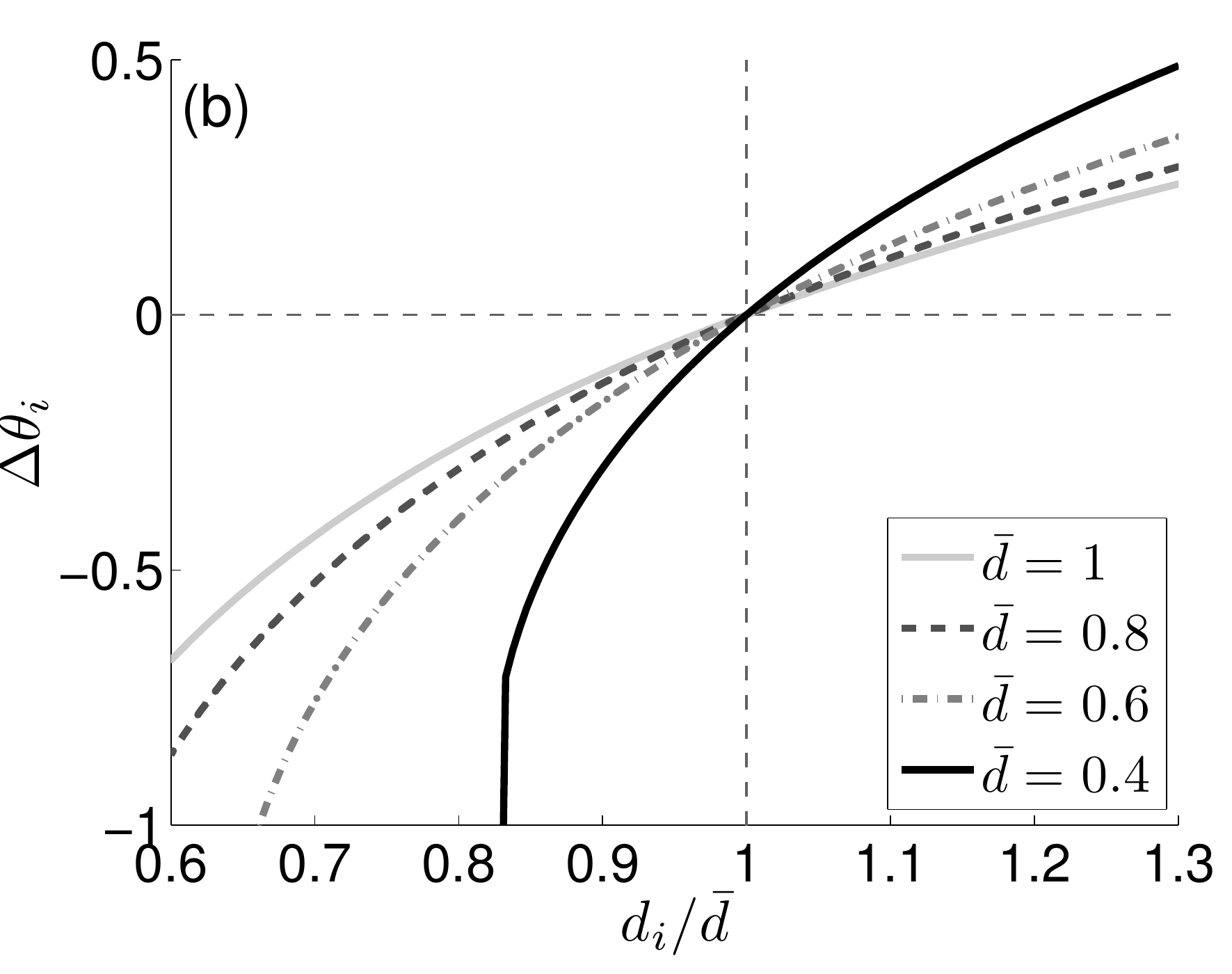}
\end{minipage} \hspace{5mm} %
\begin{minipage}[t]{0.32\textwidth}
\includegraphics[width=\textwidth]{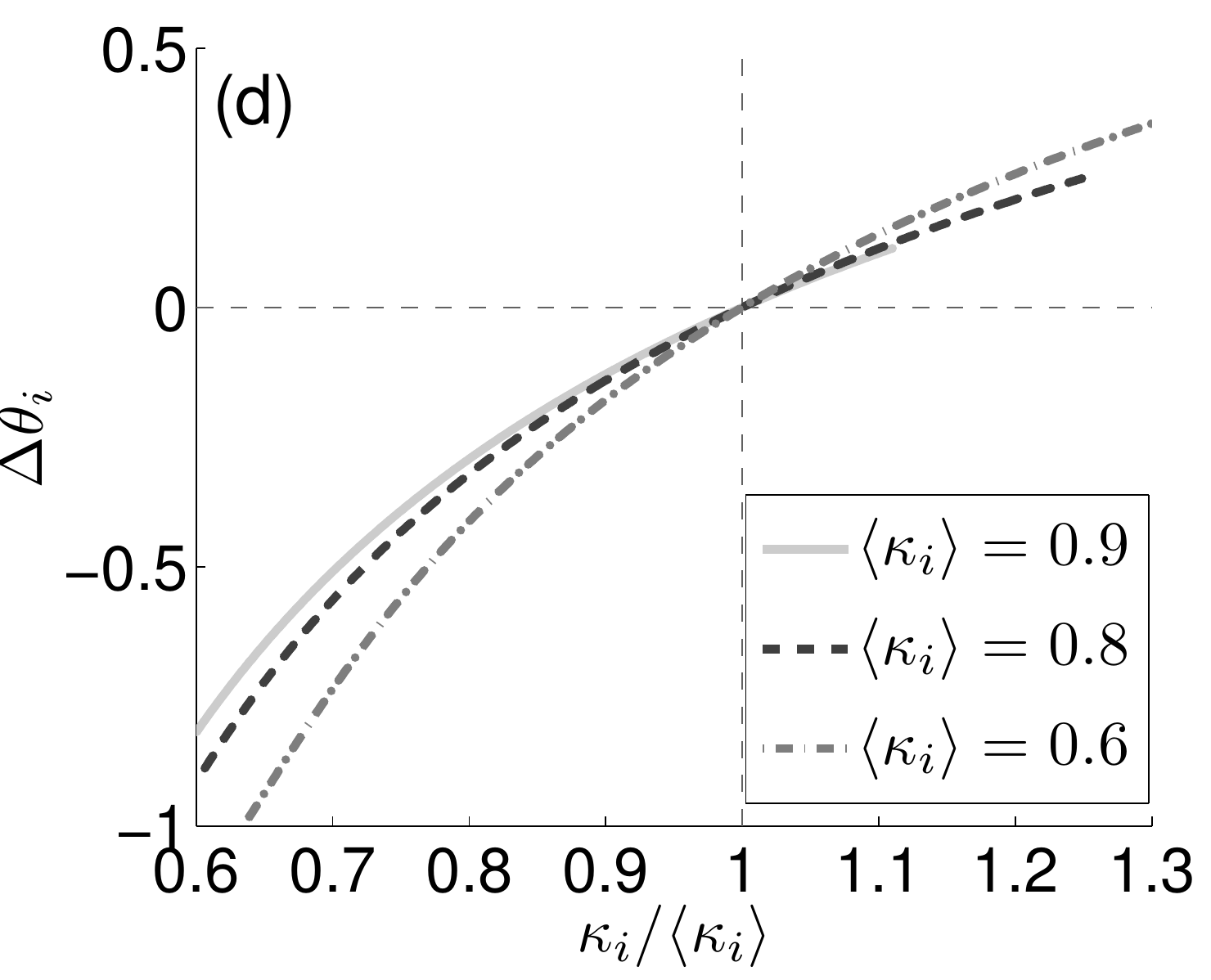}
\end{minipage}%
\caption{(a-b) Theoretical values for ratio of radius for agent $i$ to radius of swarm center of mass (a) and phase
difference (b) as a function of $d_i/\bar d$, for $\bar d = 1,\,0.8,\,0.6$ and $0.4$, with $\kappa_i = 1$. (c-d)
Theoretical values for ratio of radius for agent $i$ to radius of swarm center of mass (c) and phase difference (d) as
a function of $\kappa_i/\langle \kappa_i \rangle$, for $\langle \kappa_i \rangle = 0.9,\,0.8$ and $0.6$, for
all-to-all coupling. All calculations are done with $a = 1$ and $\tau = 4.5$ (color online). }
\label{fig:twopopkappasrot}
\end{figure*}

Fig.~\ref{fig:twopopkappasrot} (a,c) shows solution curves for $\rho_i$ and $\Delta \theta_i$
for different values of $d_i$ for a swarm where all individuals have unity acceleration factors
(thus $d_i = \mc{N}_i/N$).
Fig.~\ref{fig:twopopkappasrot} (b,d) shows solution curves for $\rho_i$ and $\Delta \theta_i$
for a swarm with all-to-all communication, where agents have heterogeneous acceleration factors.
In this case, $d_i = \kappa_i$ and $\bar d$ is the mean acceleration factor.
As shown in the figure, agents with higher acceleration factor/higher coupling have a higher radius and more positive
phase offset from the swarm center of mass than those with lower acceleration factor/lower coupling. The effects of
acceleration and coupling degree on the agents in rotation state are similar, since both factors appear in $d_i$; we
note however that variation in in the acceleration factor has a much smaller effect on the ratio
$\rho_i/\rho_{\rm rot}$ and phase offset $\Delta \theta_i$ than does breaking connections in the communication network.

\section{Finite $N$ Effects}
\label{app:finiteN}
The foregoing analysis was primarily focused on the limit of agents where $N\rightarrow \infty$. However, real networks
have finite numbers of agents; in fact, few experimental studies involve more than
a few individual agents. To explore the effectiveness of the infinite population approximation,
we conducted numerical experiments using the equations of motion \eqref{Eq:agenti} for various swarm sizes while
employing all-to-all coupling. We considered a complete graph $\mathcal{G}$ rather than an Erd{\"o}s-Renyi network
because we wish to isolate the effects of $N$ from those of incomplete connectivity. Simulations were run with random
initial conditions, i.e.\ both position and velocity were drawn from a uniform distribution with  each element of
$\bm{r}_i(0), \bm{v}_i(0) \in [0,2]$. Each set of experiments were run for 100 trials and statistics from these sets
were compared.

Using numerical simulations, we measured two quantities: a) time required to converge to the ring state, and b) the
radius of the ring state. Both quantities were measured at various population sizes, ranging from $N=2$ to $N=150$
although we only show results up to $N=100$ to focus on the small $N$ regime. We considered the system to be in the
ring state once the swarm's mean radius to its center of mass had converged to a value $R_0$, although possibly with
small fluctuations in time about it. Simulations were run for the cases of homogeneous $\kappa_i = 1$ and a uniform
distribution of $\kappa_i \in [0.2,1.0]$.

\begin{figure}
 \centering
 \begin{minipage}[t]{0.3\textwidth}
\includegraphics[width=\textwidth]{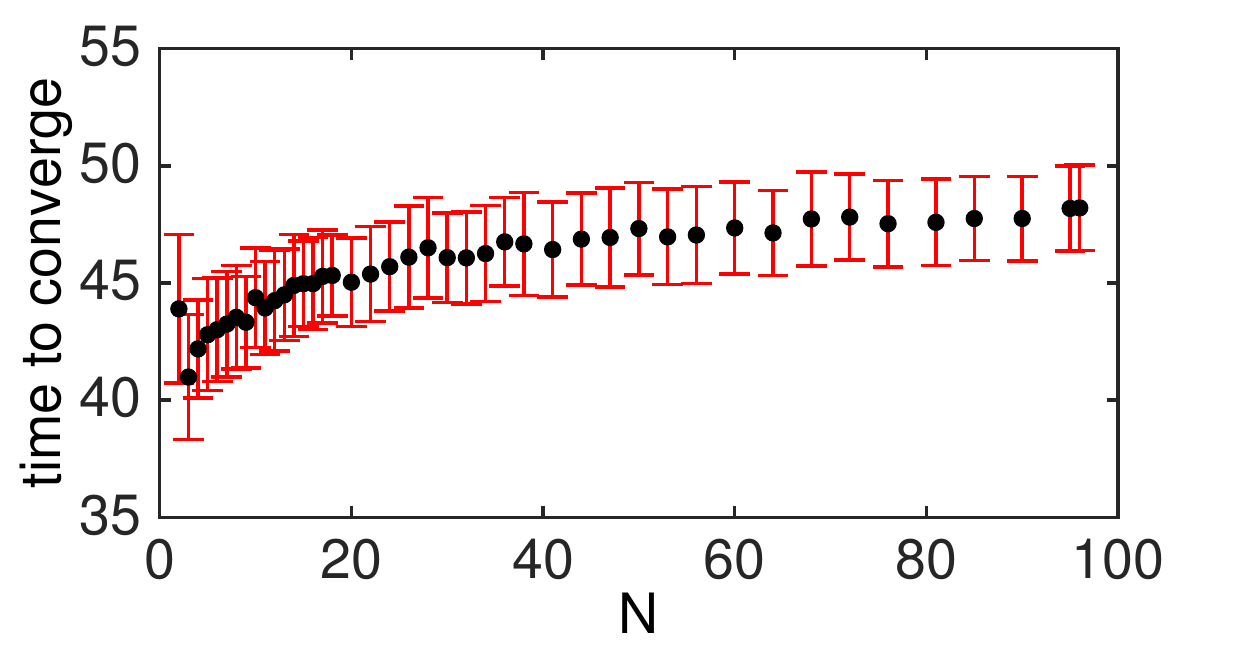}
\end{minipage}
\\
\begin{minipage}[t]{0.3\textwidth}
\includegraphics[width=\textwidth]{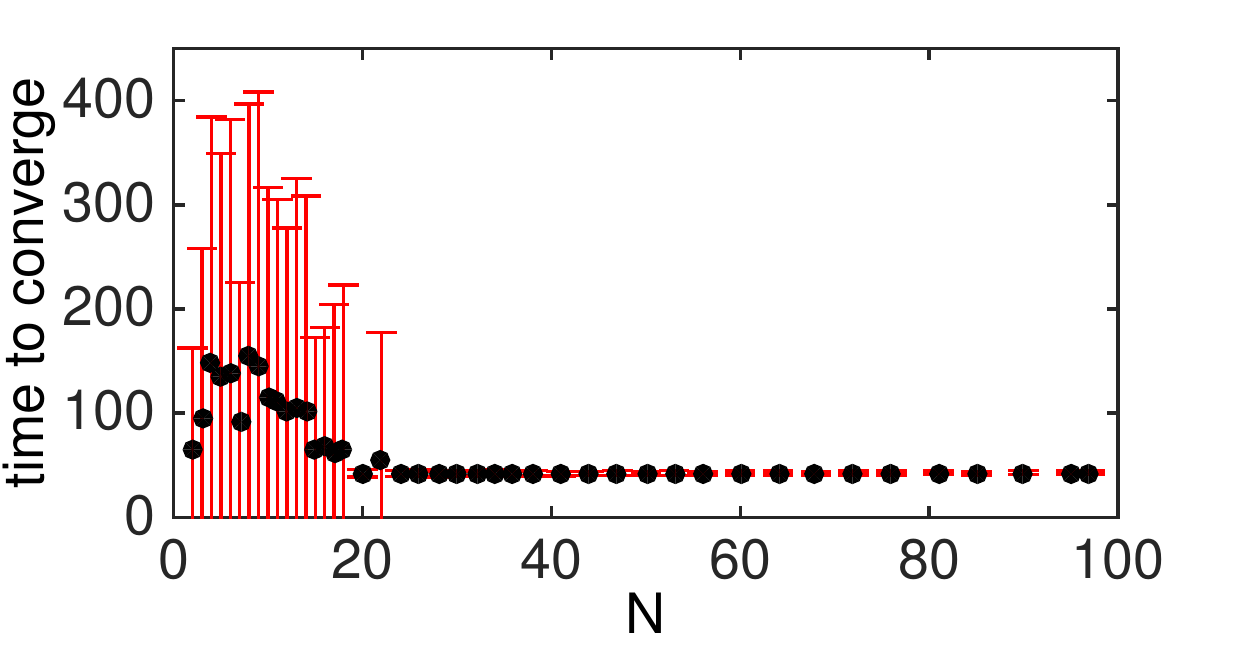}
\end{minipage}%
\caption{Time for the system to converge to ring state for different values of swarm size $N$.
Parameter values: $a = 1$, $\tau = 1$, and a) $\kappa_i = 1$ while b) $\kappa_i \in [0.2,1]$ uniformly.}
\label{fig:timetoconverge}
\end{figure}

Fig.~\ref{fig:timetoconverge}(a) shows the time to converge to the ring state for the homogeneous agent case as a
function of $N$. For large population sizes, the time to converge is relatively constant independent of $N$, but as $N$
decreases the time required and the variance of these times significantly decreases. Fig.~\ref{fig:timetoconverge}(b)
demonstrates very different behavior in the heterogeneous case; here, the time to converge is actually much greater for
smaller $N$ and converges faster for larger $N$.

When agents do converge to the ring state, we can make the following theoretical prediction for the radius $\rho$ of
the ring in the finite-$N$, $\kappa_i = 1$ case, under the assumption that all agents move the same direction along the ring:

\begin{align}
\omega^2 &= a \left(1 - \frac{1}{2}(1-\cos (\omega \tau)) \right),\\
\rho &= \frac{1}{\omega} \sqrt{1 + \frac{a \sin (\omega \tau)}{N \omega}}. \label{eq:rho_N}
\end{align}

\noindent where $\omega$ is the angular frequency of the agents moving about the ring. For $N \rightarrow \infty$ these reduce to:

\begin{align}
\omega^2 &= a,\\
\rho &=  \frac{1}{\omega},
\end{align}

\noindent which agrees with Eq.\ \eqref{eq:rho_i_limit} for the ring radius.

\begin{figure}[ht!]
\centering
\begin{minipage}[t]{0.3\textwidth}
\includegraphics[width=\textwidth]{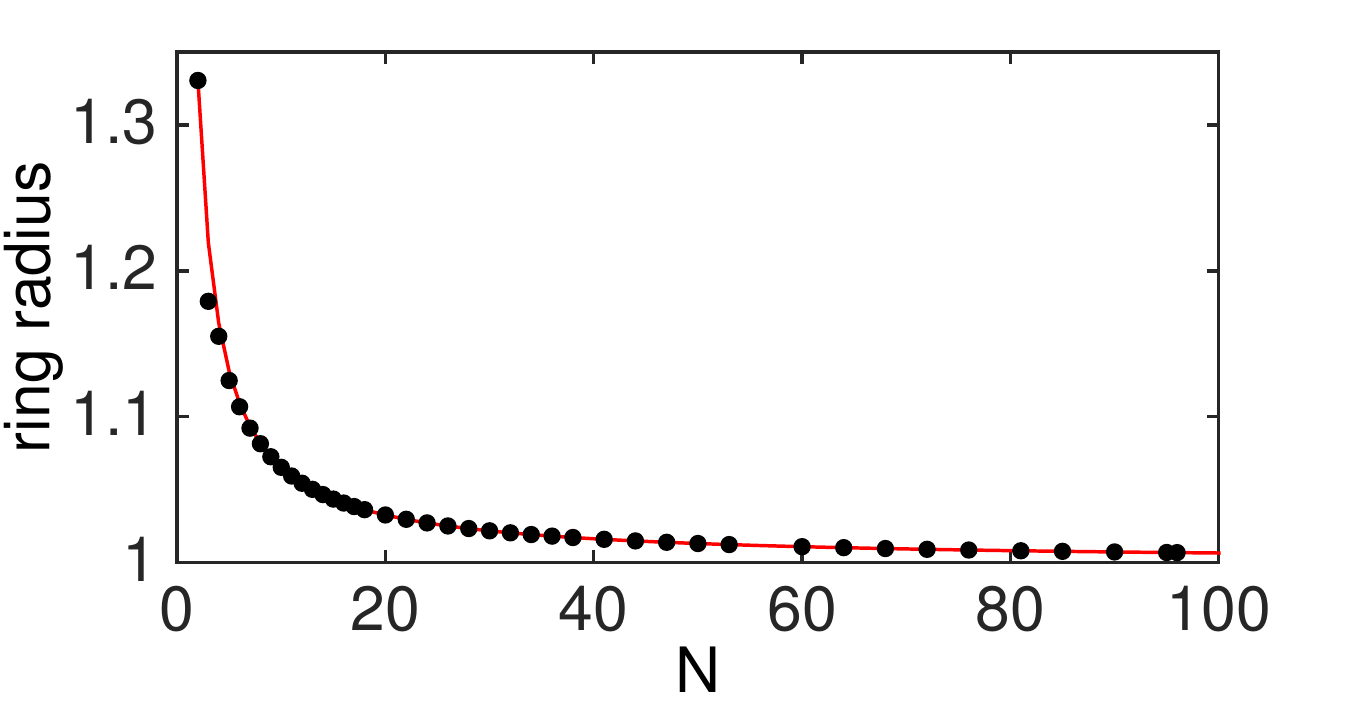}
\end{minipage} 
\\
\begin{minipage}[t]{0.3\textwidth}
\includegraphics[width=\textwidth]{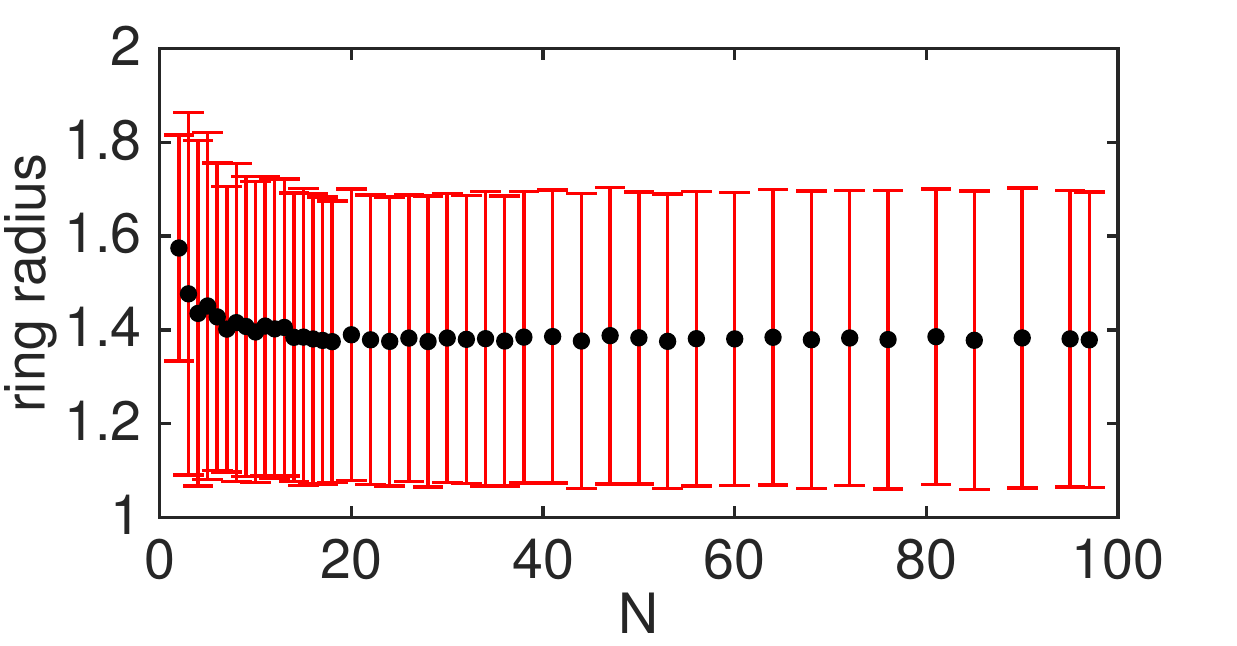}
\end{minipage}%
\caption{Radius and average speed of agents in the ring state for
different values of swarm size $N$. Black dots indicate numerical
simulations; in a) $\kappa_i = 1$ and red lines are theoretical predictions while in b)
$\kappa_i \in [0.2,1]$ uniformly and red lines show one standard deviation. Parameter
values for both cases are $a = 1$, $\tau = 1$.}
\label{fig:Radius_vs_N}
 \end{figure}

Fig.~\ref{fig:Radius_vs_N}(a) displays the radius as a function of $N$ for the homogeneous case; here, all agents
circle about the center of mass with equal radius so there is no standard deviation in the radius values, independent
of the number of agents. There is good agreement between theory in Eq.~\eqref{eq:rho_N} and stochastic simulation.
Meanwhile, Fig.~\ref{fig:Radius_vs_N}(b) shows the radius of the ring as a function of $N$ for the heterogeneous case,
where it is evident that there is a wide variation in radius, which is expected due to the uniform distribution of
acceleration factors $\kappa_i$. The effects for small $N$ in this case are similar to that of homogeneous agents; note
that the radius is expected to be larger since $\bar{\kappa} = 0.6$ resulting in a greater mean radius.

When $N$ is very small (less than 10), a wealth of new patterns emerge with what numerical simulations suggest to be
large basins of attraction. For example, when $N = 5$ the most prevalent state arranged all five agents equally along a
circle in a pentagonal pattern, rotating in the same direction. This pattern is not well described by the mean field or
the large population limit. Formally identifying these patterns and justifying their unique behavior is an area of
future work that we plan to investigate; however for the purpose of this work, the number of agents are too small to be
well-classified by an Erd{\"o}s-Renyi network.


\end{appendix}


\begin{thebibliography}{47}%
\makeatletter
\providecommand \@ifxundefined [1]{%
 \@ifx{#1\undefined}
}%
\providecommand \@ifnum [1]{%
 \ifnum #1\expandafter \@firstoftwo
 \else \expandafter \@secondoftwo
 \fi
}%
\providecommand \@ifx [1]{%
 \ifx #1\expandafter \@firstoftwo
 \else \expandafter \@secondoftwo
 \fi
}%
\providecommand \natexlab [1]{#1}%
\providecommand \enquote  [1]{``#1''}%
\providecommand \bibnamefont  [1]{#1}%
\providecommand \bibfnamefont [1]{#1}%
\providecommand \citenamefont [1]{#1}%
\providecommand \href@noop [0]{\@secondoftwo}%
\providecommand \href [0]{\begingroup \@sanitize@url \@href}%
\providecommand \@href[1]{\@@startlink{#1}\@@href}%
\providecommand \@@href[1]{\endgroup#1\@@endlink}%
\providecommand \@sanitize@url [0]{\catcode `\\12\catcode `\$12\catcode
  `\&12\catcode `\#12\catcode `\^12\catcode `\_12\catcode `\%12\relax}%
\providecommand \@@startlink[1]{}%
\providecommand \@@endlink[0]{}%
\providecommand \url  [0]{\begingroup\@sanitize@url \@url }%
\providecommand \@url [1]{\endgroup\@href {#1}{\urlprefix }}%
\providecommand \urlprefix  [0]{URL }%
\providecommand \Eprint [0]{\href }%
\providecommand \doibase [0]{http://dx.doi.org/}%
\providecommand \selectlanguage [0]{\@gobble}%
\providecommand \bibinfo  [0]{\@secondoftwo}%
\providecommand \bibfield  [0]{\@secondoftwo}%
\providecommand \translation [1]{[#1]}%
\providecommand \BibitemOpen [0]{}%
\providecommand \bibitemStop [0]{}%
\providecommand \bibitemNoStop [0]{.\EOS\space}%
\providecommand \EOS [0]{\spacefactor3000\relax}%
\providecommand \BibitemShut  [1]{\csname bibitem#1\endcsname}%
\let\auto@bib@innerbib\@empty
\bibitem [{\citenamefont {Budrene}\ and\ \citenamefont
  {Berg}(1995)}]{Budrene1995}%
  \BibitemOpen
  \bibfield  {author} {\bibinfo {author} {\bibfnamefont {E.~O.}\ \bibnamefont
  {Budrene}}\ and\ \bibinfo {author} {\bibfnamefont {H.~C.}\ \bibnamefont
  {Berg}},\ }\href@noop {} {\bibfield  {journal} {\bibinfo  {journal} {Nature}\
  }\textbf {\bibinfo {volume} {376}},\ \bibinfo {pages} {49} (\bibinfo {year}
  {1995})}\BibitemShut {NoStop}%
\bibitem [{\citenamefont {Polezhaev}\ \emph {et~al.}(2006)\citenamefont
  {Polezhaev}, \citenamefont {Pashkov}, \citenamefont {Lobanov},\ and\
  \citenamefont {Petrov}}]{Polezhaev2006}%
  \BibitemOpen
  \bibfield  {author} {\bibinfo {author} {\bibfnamefont {A.~A.}\ \bibnamefont
  {Polezhaev}}, \bibinfo {author} {\bibfnamefont {R.~A.}\ \bibnamefont
  {Pashkov}}, \bibinfo {author} {\bibfnamefont {A.~I.}\ \bibnamefont
  {Lobanov}}, \ and\ \bibinfo {author} {\bibfnamefont {I.~B.}\ \bibnamefont
  {Petrov}},\ }\href {\doibase 10.1387/ijdb.052048ap} {\bibfield  {journal}
  {\bibinfo  {journal} {The International Journal of Developmental Biology}\
  }\textbf {\bibinfo {volume} {50}},\ \bibinfo {pages} {309} (\bibinfo {year}
  {2006})}\BibitemShut {NoStop}%
\bibitem [{\citenamefont {Lee}\ \emph {et~al.}(2013)\citenamefont {Lee},
  \citenamefont {Kelley}, \citenamefont {Nordstrom}, \citenamefont
  {Ouellette},\ and\ \citenamefont {Losert}}]{Lee2013}%
  \BibitemOpen
  \bibfield  {author} {\bibinfo {author} {\bibfnamefont {R.~M.}\ \bibnamefont
  {Lee}}, \bibinfo {author} {\bibfnamefont {D.~H.}\ \bibnamefont {Kelley}},
  \bibinfo {author} {\bibfnamefont {K.~N.}\ \bibnamefont {Nordstrom}}, \bibinfo
  {author} {\bibfnamefont {N.~T.}\ \bibnamefont {Ouellette}}, \ and\ \bibinfo
  {author} {\bibfnamefont {W.}~\bibnamefont {Losert}},\ }\href {\doibase
  10.1088/1367-2630/15/2/025036} {\bibfield  {journal} {\bibinfo  {journal}
  {New Journal of Physics}\ }\textbf {\bibinfo {volume} {15}} (\bibinfo {year}
  {2013}),\ 10.1088/1367-2630/15/2/025036}\BibitemShut {NoStop}%
\bibitem [{\citenamefont {Tunstr{\o}m}\ \emph {et~al.}(2013)\citenamefont
  {Tunstr{\o}m}, \citenamefont {Katz}, \citenamefont {Ioannou}, \citenamefont
  {Huepe}, \citenamefont {Lutz},\ and\ \citenamefont {Couzin}}]{Tunstrom2013}%
  \BibitemOpen
  \bibfield  {author} {\bibinfo {author} {\bibfnamefont {K.}\ \bibnamefont
  {Tunstr{\o}m}}, \bibinfo {author} {\bibfnamefont {Y.}~\bibnamefont {Katz}},
  \bibinfo {author} {\bibfnamefont {C.~C.}\ \bibnamefont {Ioannou}}, \bibinfo
  {author} {\bibfnamefont {C.}~\bibnamefont {Huepe}}, \bibinfo {author}
  {\bibfnamefont {M.~J.}\ \bibnamefont {Lutz}}, \ and\ \bibinfo {author}
  {\bibfnamefont {I.~D.}\ \bibnamefont {Couzin}},\ }\href {\doibase
  10.1371/journal.pcbi.1002915} {\bibfield  {journal} {\bibinfo  {journal}
  {PLoS computational biology}\ }\textbf {\bibinfo {volume} {9}},\ \bibinfo
  {pages} {e1002915} (\bibinfo {year} {2013})}\BibitemShut {NoStop}%
\bibitem [{\citenamefont {Helbing}\ and\ \citenamefont
  {Molnar}(1995)}]{Helbing1995}%
  \BibitemOpen
  \bibfield  {author} {\bibinfo {author} {\bibfnamefont {D.}~\bibnamefont
  {Helbing}}\ and\ \bibinfo {author} {\bibfnamefont {P.}~\bibnamefont
  {Molnar}},\ }\href@noop {} {\bibfield  {journal} {\bibinfo  {journal}
  {Physical Review E}\ }\textbf {\bibinfo {volume} {51}},\ \bibinfo {pages}
  {4282} (\bibinfo {year} {1995})}\BibitemShut {NoStop}%
\bibitem [{\citenamefont {Lee}(2006)}]{Lee2006}%
  \BibitemOpen
  \bibfield  {author} {\bibinfo {author} {\bibfnamefont {S.-H.}\ \bibnamefont
  {Lee}},\ }\href {\doibase 10.1016/j.physleta.2006.04.065} {\bibfield
  {journal} {\bibinfo  {journal} {Physics Letters A}\ }\textbf {\bibinfo
  {volume} {357}},\ \bibinfo {pages} {270} (\bibinfo {year}
  {2006})}\BibitemShut {NoStop}%
\bibitem [{\citenamefont {Vicsek}\ \emph {et~al.}(2006)\citenamefont {Vicsek},
  \citenamefont {Czirok}, \citenamefont {Ben-Jacob}, \citenamefont {Cohen},\
  and\ \citenamefont {Shochet}}]{Vicsek2006}%
  \BibitemOpen
  \bibfield  {author} {\bibinfo {author} {\bibfnamefont {T.}~\bibnamefont
  {Vicsek}}, \bibinfo {author} {\bibfnamefont {A.}~\bibnamefont {Czirok}},
  \bibinfo {author} {\bibfnamefont {E.}~\bibnamefont {Ben-Jacob}}, \bibinfo
  {author} {\bibfnamefont {I.}~\bibnamefont {Cohen}}, \ and\ \bibinfo {author}
  {\bibfnamefont {O.}~\bibnamefont {Shochet}},\ }\href@noop {} {\enquote
  {\bibinfo {title} {{Novel type of phase transition in a system of self-driven
  particles}},}\ } (\bibinfo {year} {2006}),\ \Eprint
  {http://arxiv.org/abs/0611743v1} {arXiv:0611743v1 [arXiv:cond-mat]}
  \BibitemShut {NoStop}%
\bibitem [{\citenamefont {Edelstein-Keshet}\ \emph {et~al.}(1998)\citenamefont
  {Edelstein-Keshet}, \citenamefont {Grunbaum},\ and\ \citenamefont
  {Watmough}}]{Edelstein-Keshet1998}%
  \BibitemOpen
  \bibfield  {author} {\bibinfo {author} {\bibfnamefont {L.}~\bibnamefont
  {Edelstein-Keshet}}, \bibinfo {author} {\bibfnamefont {D.}~\bibnamefont
  {Grunbaum}}, \ and\ \bibinfo {author} {\bibfnamefont {J.}~\bibnamefont
  {Watmough}},\ }\href {\doibase 10.1007/s002850050112} {\bibfield  {journal}
  {\bibinfo  {journal} {Journal of Mathematical Biology}\ }\textbf {\bibinfo
  {volume} {36}},\ \bibinfo {pages} {515} (\bibinfo {year} {1998})}\BibitemShut
  {NoStop}%
\bibitem [{\citenamefont {Topaz}\ and\ \citenamefont
  {Bertozzi}(2004)}]{Topaz2004}%
  \BibitemOpen
  \bibfield  {author} {\bibinfo {author} {\bibfnamefont {C.~M.}\ \bibnamefont
  {Topaz}}\ and\ \bibinfo {author} {\bibfnamefont {A.~L.}\ \bibnamefont
  {Bertozzi}},\ }\href {\doibase 10.1137/S0036139903437424} {\bibfield
  {journal} {\bibinfo  {journal} {SIAM Journal on Applied Mathematics}\
  }\textbf {\bibinfo {volume} {65}},\ \bibinfo {pages} {152} (\bibinfo {year}
  {2004})}\BibitemShut {NoStop}%
\bibitem [{\citenamefont {Reynolds}(1987)}]{Reynolds1987}%
  \BibitemOpen
  \bibfield  {author} {\bibinfo {author} {\bibfnamefont {C.~W.}\ \bibnamefont
  {Reynolds}},\ }\href {\doibase 10.1145/37402.37406} {\bibfield  {journal}
  {\bibinfo  {journal} {ACM SIGGRAPH Computer Graphics}\ }\textbf {\bibinfo
  {volume} {21}},\ \bibinfo {pages} {25} (\bibinfo {year} {1987})}\BibitemShut
  {NoStop}%
\bibitem [{\citenamefont {Miller}\ \emph {et~al.}(2012)\citenamefont {Miller},
  \citenamefont {Kolpas}, \citenamefont {Neto},\ and\ \citenamefont
  {Rossi}}]{Miller2012}%
  \BibitemOpen
  \bibfield  {author} {\bibinfo {author} {\bibfnamefont {J.~M.}\ \bibnamefont
  {Miller}}, \bibinfo {author} {\bibfnamefont {A.}~\bibnamefont {Kolpas}},
  \bibinfo {author} {\bibfnamefont {J.~P.~J.}\ \bibnamefont {Neto}}, \ and\
  \bibinfo {author} {\bibfnamefont {L.~F.}\ \bibnamefont {Rossi}},\ }\href
  {\doibase 10.1007/s11538-011-9676-y} {\bibfield  {journal} {\bibinfo
  {journal} {Bulletin of mathematical biology}\ }\textbf {\bibinfo {volume}
  {74}},\ \bibinfo {pages} {536} (\bibinfo {year} {2012})}\BibitemShut
  {NoStop}%
\bibitem [{\citenamefont {Tarras}\ \emph {et~al.}(2013)\citenamefont {Tarras},
  \citenamefont {Moussa}, \citenamefont {Mazroui}, \citenamefont {Boughaleb},\
  and\ \citenamefont {Hajjaji}}]{Tarras2013}%
  \BibitemOpen
  \bibfield  {author} {\bibinfo {author} {\bibfnamefont {I.}~\bibnamefont
  {Tarras}}, \bibinfo {author} {\bibfnamefont {N.}~\bibnamefont {Moussa}},
  \bibinfo {author} {\bibfnamefont {M.}~\bibnamefont {Mazroui}}, \bibinfo
  {author} {\bibfnamefont {Y.}~\bibnamefont {Boughaleb}}, \ and\ \bibinfo
  {author} {\bibfnamefont {A.}~\bibnamefont {Hajjaji}},\ }\href {\doibase
  10.1142/S0217984913500280} {\bibfield  {journal} {\bibinfo  {journal} {Modern
  Physics Letters B}\ }\textbf {\bibinfo {volume} {27}},\ \bibinfo {pages}
  {1350028} (\bibinfo {year} {2013})}\BibitemShut {NoStop}%
\bibitem [{\citenamefont {Vir\'{a}gh}\ \emph {et~al.}(2014)\citenamefont
  {Vir\'{a}gh}, \citenamefont {V\'{a}s\'{a}rhelyi}, \citenamefont {Tarcai},
  \citenamefont {Sz\"{o}r\'{e}nyi}, \citenamefont {Somorjai}, \citenamefont
  {Nepusz},\ and\ \citenamefont {Vicsek}}]{Viragh2014}%
  \BibitemOpen
  \bibfield  {author} {\bibinfo {author} {\bibfnamefont {C.}~\bibnamefont
  {Vir\'{a}gh}}, \bibinfo {author} {\bibfnamefont {G.}~\bibnamefont
  {V\'{a}s\'{a}rhelyi}}, \bibinfo {author} {\bibfnamefont {N.}~\bibnamefont
  {Tarcai}}, \bibinfo {author} {\bibfnamefont {T.}~\bibnamefont
  {Sz\"{o}r\'{e}nyi}}, \bibinfo {author} {\bibfnamefont {G.}~\bibnamefont
  {Somorjai}}, \bibinfo {author} {\bibfnamefont {T.}~\bibnamefont {Nepusz}}, \
  and\ \bibinfo {author} {\bibfnamefont {T.}~\bibnamefont {Vicsek}},\ }\href
  {\doibase 10.1088/1748-3182/9/2/025012} {\bibfield  {journal} {\bibinfo
  {journal} {Bioinspiration \& biomimetics}\ }\textbf {\bibinfo {volume} {9}},\
  \bibinfo {pages} {025012} (\bibinfo {year} {2014})}\BibitemShut {NoStop}%
\bibitem [{\citenamefont {Vicsek}\ \emph {et~al.}(1995)\citenamefont {Vicsek},
  \citenamefont {Czirok}, \citenamefont {Ben-Jacob}, \citenamefont {Cohen},\
  and\ \citenamefont {Shochet}}]{Vicsek1995}%
  \BibitemOpen
  \bibfield  {author} {\bibinfo {author} {\bibfnamefont {T.}~\bibnamefont
  {Vicsek}}, \bibinfo {author} {\bibfnamefont {A.}~\bibnamefont {Czirok}},
  \bibinfo {author} {\bibfnamefont {E.}~\bibnamefont {Ben-Jacob}}, \bibinfo
  {author} {\bibfnamefont {I.}~\bibnamefont {Cohen}}, \ and\ \bibinfo {author}
  {\bibfnamefont {O.}~\bibnamefont {Shochet}},\ }\href@noop {} {\bibfield
  {journal} {\bibinfo  {journal} {Physical Review Letters}\ }\textbf {\bibinfo
  {volume} {75}},\ \bibinfo {pages} {1226} (\bibinfo {year}
  {1995})}\BibitemShut {NoStop}%
\bibitem [{\citenamefont {Nilsen}\ \emph {et~al.}(2013)\citenamefont {Nilsen},
  \citenamefont {Paige}, \citenamefont {Warner}, \citenamefont {Mayhew},
  \citenamefont {Sutley}, \citenamefont {Lam}, \citenamefont {Bernoff},\ and\
  \citenamefont {Topaz}}]{Nilsen2013}%
  \BibitemOpen
  \bibfield  {author} {\bibinfo {author} {\bibfnamefont {C.}~\bibnamefont
  {Nilsen}}, \bibinfo {author} {\bibfnamefont {J.}~\bibnamefont {Paige}},
  \bibinfo {author} {\bibfnamefont {O.}~\bibnamefont {Warner}}, \bibinfo
  {author} {\bibfnamefont {B.}~\bibnamefont {Mayhew}}, \bibinfo {author}
  {\bibfnamefont {R.}~\bibnamefont {Sutley}}, \bibinfo {author} {\bibfnamefont
  {M.}~\bibnamefont {Lam}}, \bibinfo {author} {\bibfnamefont {A.~J.}\
  \bibnamefont {Bernoff}}, \ and\ \bibinfo {author} {\bibfnamefont {C.~M.}\
  \bibnamefont {Topaz}},\ }\href {\doibase 10.1371/journal.pone.0083343}
  {\bibfield  {journal} {\bibinfo  {journal} {PloS ONE}\ }\textbf {\bibinfo
  {volume} {8}},\ \bibinfo {pages} {e83343} (\bibinfo {year}
  {2013})}\BibitemShut {NoStop}%
\bibitem [{\citenamefont {Ballerini}\ \emph {et~al.}(2008)\citenamefont
  {Ballerini}, \citenamefont {Cabibbo}, \citenamefont {Candelier},
  \citenamefont {Cavagna}, \citenamefont {Cisbani}, \citenamefont {Giardina},
  \citenamefont {Lecomte}, \citenamefont {Orlandi}, \citenamefont {Parisi},
  \citenamefont {Procaccini}, \citenamefont {Viale},\ and\ \citenamefont
  {Zdravkovic}}]{Ballerini2008}%
  \BibitemOpen
  \bibfield  {author} {\bibinfo {author} {\bibfnamefont {M.}~\bibnamefont
  {Ballerini}}, \bibinfo {author} {\bibfnamefont {N.}~\bibnamefont {Cabibbo}},
  \bibinfo {author} {\bibfnamefont {R.}~\bibnamefont {Candelier}}, \bibinfo
  {author} {\bibfnamefont {A.}~\bibnamefont {Cavagna}}, \bibinfo {author}
  {\bibfnamefont {E.}~\bibnamefont {Cisbani}}, \bibinfo {author} {\bibfnamefont
  {I.}~\bibnamefont {Giardina}}, \bibinfo {author} {\bibfnamefont
  {V.}~\bibnamefont {Lecomte}}, \bibinfo {author} {\bibfnamefont
  {A.}~\bibnamefont {Orlandi}}, \bibinfo {author} {\bibfnamefont
  {G.}~\bibnamefont {Parisi}}, \bibinfo {author} {\bibfnamefont
  {A.}~\bibnamefont {Procaccini}}, \bibinfo {author} {\bibfnamefont
  {M.}~\bibnamefont {Viale}}, \ and\ \bibinfo {author} {\bibfnamefont
  {V.}~\bibnamefont {Zdravkovic}},\ }\href {\doibase 10.1073/pnas.0711437105}
  {\bibfield  {journal} {\bibinfo  {journal} {Proceedings of the National
  Academy of Sciences of the United States of America}\ }\textbf {\bibinfo
  {volume} {105}},\ \bibinfo {pages} {1232} (\bibinfo {year} {2008})},\ \Eprint
  {http://arxiv.org/abs/0709.1916} {arXiv:0709.1916} \BibitemShut {NoStop}%
\bibitem [{\citenamefont {Katz}\ \emph {et~al.}(2011)\citenamefont {Katz},
  \citenamefont {Tunstr{\o}m}, \citenamefont {Ioannou}, \citenamefont {Huepe},\
  and\ \citenamefont {Couzin}}]{Katz2011}%
  \BibitemOpen
  \bibfield  {author} {\bibinfo {author} {\bibfnamefont {Y.}~\bibnamefont
  {Katz}}, \bibinfo {author} {\bibfnamefont {K.}~\bibnamefont {Tunstr{\o}m}},
  \bibinfo {author} {\bibfnamefont {C.~C.}\ \bibnamefont {Ioannou}}, \bibinfo
  {author} {\bibfnamefont {C.}~\bibnamefont {Huepe}}, \ and\ \bibinfo {author}
  {\bibfnamefont {I.~D.}\ \bibnamefont {Couzin}},\ }\href {\doibase
  10.1073/pnas.1107583108} {\bibfield  {journal} {\bibinfo  {journal}
  {Proceedings of the National Academy of Sciences}\ }\textbf {\bibinfo
  {volume} {108}},\ \bibinfo {pages} {18720} (\bibinfo {year}
  {2011})}\BibitemShut {NoStop}%
\bibitem [{\citenamefont {Calovi}\ \emph {et~al.}(2014)\citenamefont {Calovi},
  \citenamefont {Lopez}, \citenamefont {Ngo}, \citenamefont {Sire},
  \citenamefont {Chat\'{e}},\ and\ \citenamefont {Theraulaz}}]{Calovi2014}%
  \BibitemOpen
  \bibfield  {author} {\bibinfo {author} {\bibfnamefont {D.~S.}\ \bibnamefont
  {Calovi}}, \bibinfo {author} {\bibfnamefont {U.}~\bibnamefont {Lopez}},
  \bibinfo {author} {\bibfnamefont {S.}~\bibnamefont {Ngo}}, \bibinfo {author}
  {\bibfnamefont {C.}~\bibnamefont {Sire}}, \bibinfo {author} {\bibfnamefont
  {H.}~\bibnamefont {Chat\'{e}}}, \ and\ \bibinfo {author} {\bibfnamefont
  {G.}~\bibnamefont {Theraulaz}},\ }\href {\doibase
  10.1088/1367-2630/16/1/015026} {\bibfield  {journal} {\bibinfo  {journal}
  {New Journal of Physics}\ }\textbf {\bibinfo {volume} {16}} (\bibinfo {year}
  {2014}),\ 10.1088/1367-2630/16/1/015026},\ \Eprint
  {http://arxiv.org/abs/1308.2889} {arXiv:1308.2889} \BibitemShut {NoStop}%
\bibitem [{\citenamefont {Viscido}\ \emph {et~al.}(2005)\citenamefont
  {Viscido}, \citenamefont {Parrish},\ and\ \citenamefont
  {Gr\"{u}nbaum}}]{Viscido2005}%
  \BibitemOpen
  \bibfield  {author} {\bibinfo {author} {\bibfnamefont {S.~V.}\ \bibnamefont
  {Viscido}}, \bibinfo {author} {\bibfnamefont {J.~K.}\ \bibnamefont
  {Parrish}}, \ and\ \bibinfo {author} {\bibfnamefont {D.}~\bibnamefont
  {Gr\"{u}nbaum}},\ }\href {\doibase 10.1016/j.ecolmodel.2004.08.019}
  {\bibfield  {journal} {\bibinfo  {journal} {Ecological Modelling}\ }\textbf
  {\bibinfo {volume} {183}},\ \bibinfo {pages} {347} (\bibinfo {year}
  {2005})}\BibitemShut {NoStop}%
\bibitem [{\citenamefont {Martin}\ and\ \citenamefont
  {Ruan}(2001)}]{Martin2001}%
  \BibitemOpen
  \bibfield  {author} {\bibinfo {author} {\bibfnamefont {A.}~\bibnamefont
  {Martin}}\ and\ \bibinfo {author} {\bibfnamefont {S.}~\bibnamefont {Ruan}},\
  }\href {\doibase 10.1007/s002850100095} {\bibfield  {journal} {\bibinfo
  {journal} {Journal of Mathematical Biology}\ }\textbf {\bibinfo {volume}
  {43}},\ \bibinfo {pages} {247} (\bibinfo {year} {2001})}\BibitemShut
  {NoStop}%
\bibitem [{\citenamefont {Bernard}\ \emph {et~al.}(2004)\citenamefont
  {Bernard}, \citenamefont {B\'{e}lair},\ and\ \citenamefont
  {Mackey}}]{Bernard2004}%
  \BibitemOpen
  \bibfield  {author} {\bibinfo {author} {\bibfnamefont {S.}~\bibnamefont
  {Bernard}}, \bibinfo {author} {\bibfnamefont {J.}~\bibnamefont {B\'{e}lair}},
  \ and\ \bibinfo {author} {\bibfnamefont {M.~C.}\ \bibnamefont {Mackey}},\
  }\href {\doibase 10.1016/j.crvi.2003.05.005} {\bibfield  {journal} {\bibinfo
  {journal} {Comptes Rendus Biologies}\ }\textbf {\bibinfo {volume} {327}},\
  \bibinfo {pages} {201} (\bibinfo {year} {2004})}\BibitemShut {NoStop}%
\bibitem [{\citenamefont {Monk}(2003)}]{Monk2003}%
  \BibitemOpen
  \bibfield  {author} {\bibinfo {author} {\bibfnamefont {N.~A.~M.}\
  \bibnamefont {Monk}},\ }\href {\doibase 10.1016/S} {\bibfield  {journal}
  {\bibinfo  {journal} {Current Biology}\ }\textbf {\bibinfo {volume} {13}},\
  \bibinfo {pages} {1409} (\bibinfo {year} {2003})}\BibitemShut {NoStop}%
\bibitem [{\citenamefont {Giuggioli}\ \emph {et~al.}(2015)\citenamefont
  {Giuggioli}, \citenamefont {McKetterick},\ and\ \citenamefont
  {Holderied}}]{Giuggioli2015}%
  \BibitemOpen
  \bibfield  {author} {\bibinfo {author} {\bibfnamefont {L.}~\bibnamefont
  {Giuggioli}}, \bibinfo {author} {\bibfnamefont {T.~J.}\ \bibnamefont
  {McKetterick}}, \ and\ \bibinfo {author} {\bibfnamefont {M.}~\bibnamefont
  {Holderied}},\ }\href {\doibase 10.1371/journal.pcbi.1004089} {\bibfield
  {journal} {\bibinfo  {journal} {PLOS Computational Biology}\ }\textbf
  {\bibinfo {volume} {11}},\ \bibinfo {pages} {e1004089} (\bibinfo {year}
  {2015})}\BibitemShut {NoStop}%
\bibitem [{\citenamefont {Forgoston}\ and\ \citenamefont
  {Schwartz}(2008)}]{Forgoston2008}%
  \BibitemOpen
  \bibfield  {author} {\bibinfo {author} {\bibfnamefont {E.}~\bibnamefont
  {Forgoston}}\ and\ \bibinfo {author} {\bibfnamefont {I.~B.}\ \bibnamefont
  {Schwartz}},\ }\href@noop {} {\bibfield  {journal} {\bibinfo  {journal}
  {Physical Review E}\ }\textbf {\bibinfo {volume} {77}},\ \bibinfo {pages}
  {035203} (\bibinfo {year} {2008})}\BibitemShut {NoStop}%
\bibitem [{\citenamefont {{Mier-y-Teran Romero}}\ \emph
  {et~al.}(2011)\citenamefont {{Mier-y-Teran Romero}}, \citenamefont
  {Forgoston},\ and\ \citenamefont {Schwartz}}]{Romero2011}%
  \BibitemOpen
  \bibfield  {author} {\bibinfo {author} {\bibfnamefont {L.}~\bibnamefont
  {{Mier-y-Teran Romero}}}, \bibinfo {author} {\bibfnamefont {E.}~\bibnamefont
  {Forgoston}}, \ and\ \bibinfo {author} {\bibfnamefont {I.~B.}\ \bibnamefont
  {Schwartz}},\ }in\ \href {\doibase 10.1109/IROS.2011.6048160} {\emph
  {\bibinfo {booktitle} {Proceedings of the IEEE/RSJ International Conference
  on Intelligent Robots and Systems}}}\ (\bibinfo {year} {2011})\ pp.\ \bibinfo
  {pages} {3905--3910}\BibitemShut {NoStop}%
\bibitem [{\citenamefont {{Mier-y-Teran Romero}}\ \emph
  {et~al.}(2012{\natexlab{a}})\citenamefont {{Mier-y-Teran Romero}},
  \citenamefont {Forgoston},\ and\ \citenamefont {Schwartz}}]{Romero2012}%
  \BibitemOpen
  \bibfield  {author} {\bibinfo {author} {\bibfnamefont {L.}~\bibnamefont
  {{Mier-y-Teran Romero}}}, \bibinfo {author} {\bibfnamefont {E.}~\bibnamefont
  {Forgoston}}, \ and\ \bibinfo {author} {\bibfnamefont {I.~B.}\ \bibnamefont
  {Schwartz}},\ }\href {\doibase 10.1109/TRO.2012.2198511} {\bibfield
  {journal} {\bibinfo  {journal} {IEEE Transactions on Robotics}\ }\textbf
  {\bibinfo {volume} {28}},\ \bibinfo {pages} {1034} (\bibinfo {year}
  {2012}{\natexlab{a}})},\ \Eprint {http://arxiv.org/abs/arXiv:1205.0195v1}
  {arXiv:arXiv:1205.0195v1} \BibitemShut {NoStop}%
\bibitem [{\citenamefont {Lindley}\ \emph {et~al.}(2013)\citenamefont
  {Lindley}, \citenamefont {{Mier-y-Teran Romero}},\ and\ \citenamefont
  {Schwartz}}]{Lindley2013a}%
  \BibitemOpen
  \bibfield  {author} {\bibinfo {author} {\bibfnamefont {B.~S.}\ \bibnamefont
  {Lindley}}, \bibinfo {author} {\bibfnamefont {L.}~\bibnamefont {{Mier-y-Teran
  Romero}}}, \ and\ \bibinfo {author} {\bibfnamefont {I.~B.}\ \bibnamefont
  {Schwartz}},\ }in\ \href {\doibase 10.1109/ACC.2013.6580546} {\emph {\bibinfo
  {booktitle} {2013 American Control Conference}}}\ (\bibinfo  {publisher}
  {Ieee},\ \bibinfo {year} {2013})\ pp.\ \bibinfo {pages}
  {4587--4591}\BibitemShut {NoStop}%
\bibitem [{\citenamefont {Liu}\ \emph {et~al.}(2003)\citenamefont {Liu},
  \citenamefont {Passino},\ and\ \citenamefont {Polycarpou}}]{Liu2003}%
  \BibitemOpen
  \bibfield  {author} {\bibinfo {author} {\bibfnamefont {Y.}~\bibnamefont
  {Liu}}, \bibinfo {author} {\bibfnamefont {K.~M.}\ \bibnamefont {Passino}}, \
  and\ \bibinfo {author} {\bibfnamefont {M.}~\bibnamefont {Polycarpou}},\
  }\href {\doibase 10.1109/TAC.2003.817942} {\bibfield  {journal} {\bibinfo
  {journal} {IEEE Transactions on Automatic Control}\ }\textbf {\bibinfo
  {volume} {48}},\ \bibinfo {pages} {1848} (\bibinfo {year}
  {2003})}\BibitemShut {NoStop}%
\bibitem [{\citenamefont {Motsch}\ and\ \citenamefont
  {Tadmor}(2011)}]{Motsch2011}%
  \BibitemOpen
  \bibfield  {author} {\bibinfo {author} {\bibfnamefont {S.}~\bibnamefont
  {Motsch}}\ and\ \bibinfo {author} {\bibfnamefont {E.}~\bibnamefont
  {Tadmor}},\ }\href {\doibase 10.1007/s10955-011-0285-9} {\bibfield  {journal}
  {\bibinfo  {journal} {Journal of Statistical Physics}\ }\textbf {\bibinfo
  {volume} {144}},\ \bibinfo {pages} {923} (\bibinfo {year}
  {2011})}\BibitemShut {NoStop}%
\bibitem [{\citenamefont {Chen}\ \emph {et~al.}(2011)\citenamefont {Chen},
  \citenamefont {Liao},\ and\ \citenamefont {Chu}}]{Chen2011}%
  \BibitemOpen
  \bibfield  {author} {\bibinfo {author} {\bibfnamefont {Z.}~\bibnamefont
  {Chen}}, \bibinfo {author} {\bibfnamefont {H.}~\bibnamefont {Liao}}, \ and\
  \bibinfo {author} {\bibfnamefont {T.}~\bibnamefont {Chu}},\ }\href {\doibase
  10.1209/0295-5075/96/40015} {\bibfield  {journal} {\bibinfo  {journal} {EPL
  (Europhysics Letters)}\ }\textbf {\bibinfo {volume} {96}},\ \bibinfo {pages}
  {40015} (\bibinfo {year} {2011})}\BibitemShut {NoStop}%
\bibitem [{\citenamefont {Chen}\ and\ \citenamefont
  {Kolokolnikov}(2014)}]{Chen2014}%
  \BibitemOpen
  \bibfield  {author} {\bibinfo {author} {\bibfnamefont {Y.}~\bibnamefont
  {Chen}}\ and\ \bibinfo {author} {\bibfnamefont {T.}~\bibnamefont
  {Kolokolnikov}},\ }\href {\doibase 10.1098/rsif.2013.1208} {\bibfield
  {journal} {\bibinfo  {journal} {Journal of the Royal Society, Interface}\
  }\textbf {\bibinfo {volume} {11}},\ \bibinfo {pages} {20131208} (\bibinfo
  {year} {2014})}\BibitemShut {NoStop}%
\bibitem [{\citenamefont {Vecil}\ \emph {et~al.}(2013)\citenamefont {Vecil},
  \citenamefont {Lafitte},\ and\ \citenamefont {{Rosado
  Linares}}}]{Vecil2013a}%
  \BibitemOpen
  \bibfield  {author} {\bibinfo {author} {\bibfnamefont {F.}~\bibnamefont
  {Vecil}}, \bibinfo {author} {\bibfnamefont {P.}~\bibnamefont {Lafitte}}, \
  and\ \bibinfo {author} {\bibfnamefont {J.}~\bibnamefont {{Rosado Linares}}},\
  }\href {\doibase 10.1016/j.physd.2012.12.010} {\bibfield  {journal} {\bibinfo
   {journal} {Physica D}\ }\textbf {\bibinfo {volume} {260}},\ \bibinfo {pages}
  {127} (\bibinfo {year} {2013})}\BibitemShut {NoStop}%
\bibitem [{\citenamefont {von Brecht}\ \emph {et~al.}(2013)\citenamefont {von
  Brecht}, \citenamefont {Kolokolnikov}, \citenamefont {Bertozzi},\ and\
  \citenamefont {Sun}}]{VonBrecht2013}%
  \BibitemOpen
  \bibfield  {author} {\bibinfo {author} {\bibfnamefont {J.~H.}\ \bibnamefont
  {von Brecht}}, \bibinfo {author} {\bibfnamefont {T.}~\bibnamefont
  {Kolokolnikov}}, \bibinfo {author} {\bibfnamefont {A.~L.}\ \bibnamefont
  {Bertozzi}}, \ and\ \bibinfo {author} {\bibfnamefont {H.}~\bibnamefont
  {Sun}},\ }\href {\doibase 10.1007/s10955-012-0680-x} {\bibfield  {journal}
  {\bibinfo  {journal} {Journal of Statistical Physics}\ }\textbf {\bibinfo
  {volume} {151}},\ \bibinfo {pages} {150} (\bibinfo {year}
  {2013})}\BibitemShut {NoStop}%
\bibitem [{\citenamefont {Steinberg}(1963)}]{Steinberg1963}%
  \BibitemOpen
  \bibfield  {author} {\bibinfo {author} {\bibfnamefont {M.~S.}\ \bibnamefont
  {Steinberg}},\ }\href {\doibase 10.1126/science.141.3579.401} {\bibfield
  {journal} {\bibinfo  {journal} {Science}\ }\textbf {\bibinfo {volume}
  {141}},\ \bibinfo {pages} {401} (\bibinfo {year} {1963})}\BibitemShut
  {NoStop}%
\bibitem [{\citenamefont {Graner}(1993)}]{Graner1993}%
  \BibitemOpen
  \bibfield  {author} {\bibinfo {author} {\bibfnamefont {F.}~\bibnamefont
  {Graner}},\ }\href@noop {} {\bibfield  {journal} {\bibinfo  {journal}
  {Physical Review E}\ }\textbf {\bibinfo {volume} {47}},\ \bibinfo {pages}
  {2128} (\bibinfo {year} {1993})}\BibitemShut {NoStop}%
\bibitem [{\citenamefont {{Mier-y-Teran Romero}}\ \emph
  {et~al.}(2012{\natexlab{b}})\citenamefont {{Mier-y-Teran Romero}},
  \citenamefont {Lindley},\ and\ \citenamefont
  {Schwartz}}]{Mier-y-Teran-Romero2012a}%
  \BibitemOpen
  \bibfield  {author} {\bibinfo {author} {\bibfnamefont {L.}~\bibnamefont
  {{Mier-y-Teran Romero}}}, \bibinfo {author} {\bibfnamefont {B.}~\bibnamefont
  {Lindley}}, \ and\ \bibinfo {author} {\bibfnamefont {I.~B.}\ \bibnamefont
  {Schwartz}},\ }\href {\doibase 10.1103/PhysRevE.86.056202} {\bibfield
  {journal} {\bibinfo  {journal} {Physical Review E}\ }\textbf {\bibinfo
  {volume} {86}},\ \bibinfo {pages} {056202} (\bibinfo {year}
  {2012}{\natexlab{b}})}\BibitemShut {NoStop}%
\bibitem [{\citenamefont {{Mier-y-Teran Romero}}\ and\ \citenamefont
  {Schwartz}(2014)}]{Mier-y-Teran-Romero2014}%
  \BibitemOpen
  \bibfield  {author} {\bibinfo {author} {\bibfnamefont {L.}~\bibnamefont
  {{Mier-y-Teran Romero}}}\ and\ \bibinfo {author} {\bibfnamefont {I.~B.}\
  \bibnamefont {Schwartz}},\ }\href {\doibase 10.1209/0295-5075/105/20002}
  {\bibfield  {journal} {\bibinfo  {journal} {EPL (Europhysics Letters)}\
  }\textbf {\bibinfo {volume} {105}},\ \bibinfo {pages} {20002} (\bibinfo
  {year} {2014})}\BibitemShut {NoStop}%
\bibitem [{\citenamefont {Leverentz}\ \emph {et~al.}(2009)\citenamefont
  {Leverentz}, \citenamefont {Topaz},\ and\ \citenamefont
  {Bernoff}}]{Leverentz2009}%
  \BibitemOpen
  \bibfield  {author} {\bibinfo {author} {\bibfnamefont {A.~J.}\ \bibnamefont
  {Leverentz}}, \bibinfo {author} {\bibfnamefont {C.~M.}\ \bibnamefont
  {Topaz}}, \ and\ \bibinfo {author} {\bibfnamefont {A.~J.}\ \bibnamefont
  {Bernoff}},\ }\href {\doibase 10.1137/090749037} {\bibfield  {journal}
  {\bibinfo  {journal} {SIAM Journal on Applied Dynamical Systems}\ }\textbf
  {\bibinfo {volume} {8}},\ \bibinfo {pages} {880} (\bibinfo {year}
  {2009})}\BibitemShut {NoStop}%
\bibitem [{\citenamefont {Burger}\ \emph {et~al.}(2013)\citenamefont {Burger},
  \citenamefont {Ha\v{s}kovec},\ and\ \citenamefont {Wolfram}}]{Burger2013}%
  \BibitemOpen
  \bibfield  {author} {\bibinfo {author} {\bibfnamefont {M.}~\bibnamefont
  {Burger}}, \bibinfo {author} {\bibfnamefont {J.}~\bibnamefont
  {Ha\v{s}kovec}}, \ and\ \bibinfo {author} {\bibfnamefont {M.-T.}\
  \bibnamefont {Wolfram}},\ }\href {\doibase 10.1016/j.physd.2012.11.003}
  {\bibfield  {journal} {\bibinfo  {journal} {Physica D. Nonlinear phenomena}\
  }\textbf {\bibinfo {volume} {260}},\ \bibinfo {pages} {145} (\bibinfo {year}
  {2013})}\BibitemShut {NoStop}%
\bibitem [{\citenamefont {Szwaykowska}\ \emph
  {et~al.}(2015{\natexlab{a}})\citenamefont {Szwaykowska}, \citenamefont
  {Romero},\ and\ \citenamefont {Schwartz}}]{Szwaykowska2014}%
  \BibitemOpen
  \bibfield  {author} {\bibinfo {author} {\bibfnamefont {K.}~\bibnamefont
  {Szwaykowska}}, \bibinfo {author} {\bibfnamefont {L.}~\bibnamefont
  {Mier-y-Teran Romero}}, \ and\ \bibinfo {author} {\bibfnamefont {I.~B.}\ \bibnamefont
  {Schwartz}},\ }\href {\doibase 10.1109/TASE.2015.2403253} {\bibfield
  {journal} {\bibinfo  {journal} {IEEE Transactions on Automation Science and
  Engineering}\ }\textbf {\bibinfo {volume} {12}},\ \bibinfo {pages} {810}
  (\bibinfo {year} {2015}{\natexlab{a}})},\ \Eprint
  {http://arxiv.org/abs/1409.1042} {arXiv:1409.1042} \BibitemShut {NoStop}%
\bibitem [{\citenamefont {Day}\ \emph {et~al.}(2015)\citenamefont {Day},
  \citenamefont {Clement}, \citenamefont {Russo}, \citenamefont {Davis},\ and\
  \citenamefont {Chung}}]{Day2015}%
  \BibitemOpen
  \bibfield  {author} {\bibinfo {author} {\bibfnamefont {M.~A.}\ \bibnamefont
  {Day}}, \bibinfo {author} {\bibfnamefont {M.~R.}\ \bibnamefont {Clement}},
  \bibinfo {author} {\bibfnamefont {J.~D.}\ \bibnamefont {Russo}}, \bibinfo
  {author} {\bibfnamefont {D.}~\bibnamefont {Davis}}, \ and\ \bibinfo {author}
  {\bibfnamefont {T.~H.}\ \bibnamefont {Chung}},\ }in\ \href@noop {} {\emph
  {\bibinfo {booktitle} {International Conference on Unmanned Aircraft Systems
  (ICUAS)}}}\ (\bibinfo {year} {2015})\ pp.\ \bibinfo {pages}
  {426--435}\BibitemShut {NoStop}%
\bibitem [{\citenamefont {Jia}\ and\ \citenamefont {Wang}(2014)}]{Jia2014}%
  \BibitemOpen
  \bibfield  {author} {\bibinfo {author} {\bibfnamefont {Y.}~\bibnamefont
  {Jia}}\ and\ \bibinfo {author} {\bibfnamefont {L.}~\bibnamefont {Wang}},\
  }\href {\doibase 10.1016/j.conengprac.2014.05.004} {\bibfield  {journal}
  {\bibinfo  {journal} {Control Engineering Practice}\ }\textbf {\bibinfo
  {volume} {30}},\ \bibinfo {pages} {1} (\bibinfo {year} {2014})}\BibitemShut
  {NoStop}%
\bibitem [{\citenamefont {Chen}\ \emph {et~al.}(2014)\citenamefont {Chen},
  \citenamefont {Zhang}, \citenamefont {Fan}, \citenamefont {Wang},\ and\
  \citenamefont {Li}}]{Chen2014b}%
  \BibitemOpen
  \bibfield  {author} {\bibinfo {author} {\bibfnamefont {Z.}~\bibnamefont
  {Chen}}, \bibinfo {author} {\bibfnamefont {H.-T.}\ \bibnamefont {Zhang}},
  \bibinfo {author} {\bibfnamefont {M.-C.}\ \bibnamefont {Fan}}, \bibinfo
  {author} {\bibfnamefont {D.}~\bibnamefont {Wang}}, \ and\ \bibinfo {author}
  {\bibfnamefont {D.}~\bibnamefont {Li}},\ }\href@noop {} {\bibfield  {journal}
  {\bibinfo  {journal} {IEEE Transactions on Control Systems Technology}\
  }\textbf {\bibinfo {volume} {22}},\ \bibinfo {pages} {1544} (\bibinfo {year}
  {2014})}\BibitemShut {NoStop}%
\bibitem [{\citenamefont {Rubenstein}\ \emph {et~al.}(2014)\citenamefont
  {Rubenstein}, \citenamefont {Cornejo},\ and\ \citenamefont
  {Nagpal}}]{Rubenstein2014}%
  \BibitemOpen
  \bibfield  {author} {\bibinfo {author} {\bibfnamefont {M.}~\bibnamefont
  {Rubenstein}}, \bibinfo {author} {\bibfnamefont {A.}~\bibnamefont {Cornejo}},
  \ and\ \bibinfo {author} {\bibfnamefont {R.}~\bibnamefont {Nagpal}},\ }\href
  {\doibase 10.1126/science.1254295} {\bibfield  {journal} {\bibinfo  {journal}
  {Science}\ }\textbf {\bibinfo {volume} {345}},\ \bibinfo {pages} {795}
  (\bibinfo {year} {2014})}\BibitemShut {NoStop}%
\bibitem [{\citenamefont {Szwaykowska}\ \emph
  {et~al.}(2015{\natexlab{b}})\citenamefont {Szwaykowska}, \citenamefont
  {Romero},\ and\ \citenamefont {Schwartz}}]{Szwaykowska2015}%
  \BibitemOpen
  \bibfield  {author} {\bibinfo {author} {\bibfnamefont {K.}~\bibnamefont
  {Szwaykowska}}, \bibinfo {author} {\bibfnamefont {L.}~\bibnamefont
  {Mier-y-Teran Romero}}, \ and\ \bibinfo {author} {\bibfnamefont {I.~B.}\ \bibnamefont
  {Schwartz}},\ }in\ \href@noop {} {\emph {\bibinfo {booktitle} {Conference on
  Decision and Control (accepted)}}}\ (\bibinfo {year} {2015})\BibitemShut
  {NoStop}%
\bibitem [{\citenamefont {Mikhailov}\ and\ \citenamefont
  {Zanette}(1999)}]{Mikhailov1999}%
  \BibitemOpen
  \bibfield  {author} {\bibinfo {author} {\bibfnamefont {A.~S.}\ \bibnamefont
  {Mikhailov}}\ and\ \bibinfo {author} {\bibfnamefont {D.~D.~H.}\ \bibnamefont
  {Zanette}},\ }\href {\doibase 10.1103/PhysRevE.60.4571} {\bibfield  {journal}
  {\bibinfo  {journal} {Physical Review E}\ }\textbf {\bibinfo {volume} {60}},\
  \bibinfo {pages} {4571} (\bibinfo {year} {1999})},\ \Eprint
  {http://arxiv.org/abs/9906004v1} {arXiv:9906004v1 [arXiv:adap-org]}
  \BibitemShut {NoStop}%
\bibitem [{\citenamefont {Erdmann}\ \emph {et~al.}(2004)\citenamefont
  {Erdmann}, \citenamefont {Ebeling},\ and\ \citenamefont
  {Mikhailov}}]{Erdmann2005}%
  \BibitemOpen
  \bibfield  {author} {\bibinfo {author} {\bibfnamefont {U.}~\bibnamefont
  {Erdmann}}, \bibinfo {author} {\bibfnamefont {W.}~\bibnamefont {Ebeling}}, \
  and\ \bibinfo {author} {\bibfnamefont {A.~S.}\ \bibnamefont {Mikhailov}},\
  }\href {\doibase 10.1103/PhysRevE.71.051904} {\bibfield  {journal} {\bibinfo
  {journal} {Physical Review E}\ }\textbf {\bibinfo {volume} {71}},\ \bibinfo
  {pages} {1} (\bibinfo {year} {2004})},\ \Eprint
  {http://arxiv.org/abs/0412037} {arXiv:0412037 [physics]} \BibitemShut
  {NoStop}%
\end{thebibliography}
\end{document}